\documentclass[%
 aip,
rsi,%
amsmath,amssymb,
preprint,%
]{revtex4-1}

\usepackage{graphicx}
\usepackage{subfig}
\usepackage{amssymb}
\usepackage{float}
\usepackage{todonotes}
\usepackage{changes}
\usepackage{color}
\usepackage{todonotes}
\usepackage{gensymb}

\usepackage{amsmath}
\usepackage{cancel}

\usepackage{algpseudocode,caption}
\usepackage[english]{babel}
\usepackage[utf8]{inputenc}
\usepackage{algorithm}

\def\ee{{\rm e}}
\def\ii{{\rm i}}

\usepackage{bm}
\def\ii{{\rm i}}

\begin{document}

\preprint{AIP/123-QED}
\title
{An inverse technique for reconstructing ocean’s density stratification from surface data}
\author{Subhajit Kar}


\author{Anirban Guha}
 \email{anirbanguha.ubc@gmail.com}
\affiliation{
Environmental and Geophysical Fluids Group, Department of Mechanical Engineering, Indian Institute of Technology Kanpur, U.P. 208016, India.\\}%

\date{\today}

\begin{abstract}
In this article, we propose an inverse technique that accurately reconstructs the ocean's density stratification profile simply from free surface elevation data. Satellite observations suggest that ocean surface contains the signature of internal tides, which are internal gravity waves generated by the barotropic tides. Since internal tides contain the information of ocean's density stratification, the latter can in principle be reconstructed from the free surface signature. We consider a simple theoretical model that approximates a continuously stratified ocean as discrete layers of constant buoyancy frequency; this facilitates the derivation of a closed-form dispersion relation. First, we numerically simulate internal tide generation for toy ocean scenarios and subsequently perform Space-Time Fourier Transform (STFT) of the free surface, which yields internal tide spectra with wavenumbers corresponding to the tidal frequency.  The density profile is reconstructed by substituting these wavenumbers into the dispersion relation. Finally, we consider a more realistic situation with rotation, bottom topography, shear and density profiles representative of the Strait of Gibraltar. Density reconstruction in the presence and absence of shear are respectively found to be $90.2\%$ and $94.2\%$ accurate. 
\end{abstract}



\maketitle
\section{Introduction}
The oceans are by and large stably stratified, that is, the density of ocean monotonically increases with depth. The ocean's density also varies with latitude and longitude,  as well as with seasons. Depending upon the strength of  
stratification, in general the vertical structure of the
ocean's density is divided into three major layers: (i) \emph{top} - weakly stratified surface mixed layer, (ii) \emph{middle} - strongly stratified pycnocline, and  (iii) \emph{bottom} - weakly stratified abyss\cite{sutherland2010internal}. 

An accurate knowledge of the ocean's density field is crucial for ocean and climate  modeling \cite{cummins1991deep}. The oceanic density stratification  also has a direct impact on the  aquatic ecosystem. In oceans and lakes, microbiological activities and accumulation of organisms are strongly affected by the pycnocline \cite{doostmohammadi2012low}. The density stratification influences the formation of spring phytoplankton blooms, which in turns help to maintain a balanced ecosystem \cite{sherman1998transitions}. 
{In particular, depth of the top mixed layer modulates the interaction between the light availability for photosynthesis and the nutrient supply from the deep oceans \cite{capotondi2012enhanced}. The density gradient at the base of the mixed layer affects the entrainment process, which plays an important role in mixed layer deepening and in supplying nutrients to the photic zone \cite{capotondi2012enhanced}. }

The ocean's density is a function of  both temperature and salinity, both of which are measured using CTD (Conductivity, Temperature and Depth) sensors using the ARGO floats \cite{bradshaw1980electrical}. These sensors, while descending (or ascending) through the ocean water, collect the necessary information.  The vertical profiles of temperature and salinity thus obtained are then substituted into the equation of state to yield ocean's density profile at a given latitude--longitude. 
At present, there is a global array of $\sim 3800$ free-drifting ARGO floats in the global ocean. Indeed, these drifters do provide an accurate measurement of the density field, however, they can not provide spatially uniform resolution data. Another drawback of this measurement technique is that these floats behave as free-drifters, therefore, their measurement at a particular point of interest cannot be controlled precisely.

The above drawbacks have persuaded us to look into a useful alternative measurement technique. To the best of our knowledge, there are no indirect or `non-invasive' techniques that can estimate the oceanic density profile. 
In this article, we propose a strategy that can provide a reasonably accurate estimate of ocean's density stratification profile (and hence, the pycnocline depth) in a fully non-invasive manner by \emph{only} analyzing the ocean free surface. To achieve this, we scrutinize one of the most important consequences of ocean's stable density stratification -- the internal tides, which are internal gravity waves (IGWs) forced at the tidal frequency \cite{wunsch1975internal}. The proposed method provides an optimal compromise between fidelity and simplicity of representation of the ocean density field.
To analyse the ocean surface imprint, we invoke a simpler approach by discretizing the vertical variation of density into discrete layers with a linear variation of density (implying the buoyancy frequency is piecewise constant) and construct a theoretical model to estimate the layered density profile.
{Recently, internal tides have been proposed as a cost-effective tool to infer the change in the upper ocean temperature due to a change in travel time of the low-mode internal tides\cite{zhao2016internal}.} 
Apart from the internal tides, there is another type of IGW that are  generated by the wind-driven flow and have been observed as a prominent peak in the Garrett \& Munk continuous internal waves spectrum \cite{alford2016near}. These wind-induced internal waves generated in the ocean mixed layer are commonly known as ``near-inertial waves'', and as the name suggests, the frequency of these waves is very close to the \emph{Coriolis} frequency \cite{d1989decay,garrett2001near}. These waves predominantly undergo downward propagation \cite{garrett2001near} and are different from the kind of IGWs the current work is fully based on  -- the \emph{internal tides}. From here on, the acronym ``IGW'' would only represent internal gravity waves generated by the semi-diurnal tides.

\begin{figure*}
\centering
\includegraphics[width=1.0\linewidth]{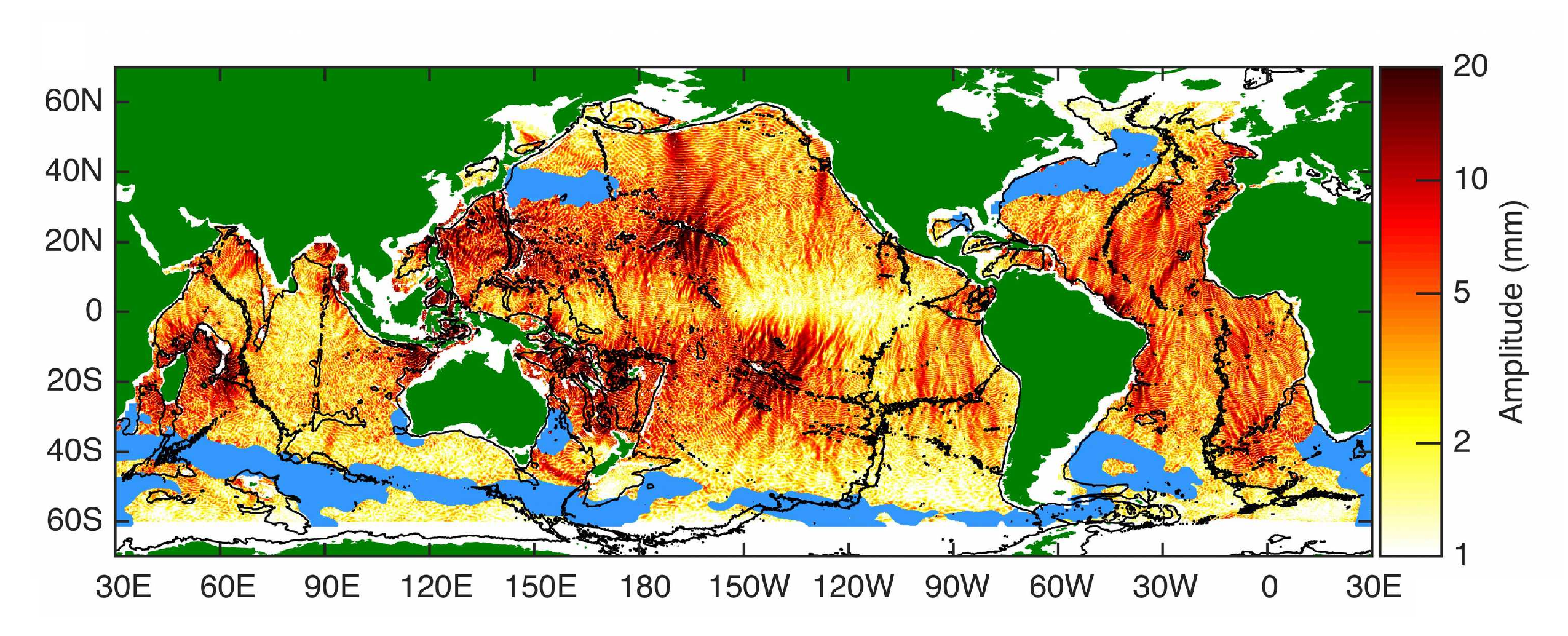}
\caption{Sea surface height amplitude of mode-$1$ semi-diurinal tides ($M_2$) from multi-satellite altimetry. The $3000$ m isobath contours are shown in black. The light blue masks show regions of the high meso-scale eddies. The image has been taken from  \citet{zhao2016global}. $\copyright$ American Meteorological Society. Used with permission.}
\label{fig:zhao_jpo}
\end{figure*}

The high-mode IGWs often dissipate near their generation site. On the contrary, the low-modes  generally travel hundreds and even a thousand kilometers before getting dissipated \cite{zhao2016global,st2002role}. 
An important aspect of low-mode IGWs is that they are very efficient in transporting momentum and energy over large distances, and help in mixing nutrients, oxygen and heat in the oceans \cite{sarkar2017topographic}. 
Turbulence and mixing due to IGW breaking play an important role in regulating the global oceanic circulation, and are one of the major factors in the climate-forecast models \cite{ferrari2009ocean}. 

The signature of the low modes on the ocean surface are detectable by the satellite altimeters \cite{ray1996surface,ray2011non}; they appear as wave-like perturbations having a frequency of the tidal frequency. { In the past two decades, efforts have been made to construct the coherent structure of the stationary low-mode internal-tides using $20$ years of sea-surface height (SSH) data  from multiple satellite altimeters \cite{zhao2012mapping, zhao2016global}. 
Fig. \ref{fig:zhao_jpo} shows the global estimation of mode-$1$ SSH using  multi-satellite altimetry. The figure also shows the regions of  strong internal tide generation sites, which are highly correlated with the large-scale topographic features, shown by the solid black lines.
The sampling rate of satellite altimeter is usually larger than the tidal  period, but remarkably enough,  IGWs can \emph{still} be studied with a dataset exhibiting sampling only once in every $20$--$60$ tidal periods \cite{zhao2012mapping}.} An astounding point comes from these observations that the signature of the IGWs remain both spatially and temporary coherent despite the fact that they are often contaminated by the meso-scale eddies or the sub-surface shear (mostly induced by the wind forcing) \cite{zhao2016global}. 
The pivotal point of our article is the realization that the ocean surface signature of IGWs (which, as already mentioned, are well detectable via satellite observations) carry the information of ocean's density stratification, and can in principle be inverted to reconstruct the latter.

We have organized the article as follows. In Sec. \ref{sec:1} we discuss the closed form dispersion relations for one, two and three-layered of constant buoyancy frequency. 
{Additionally, we have provided a  mathematical justification regarding the uniqueness of  IGW wavenumbers, and have also highlighted a unique situation where two different density profiles provide the exact same set of wavenumbers.}
 {Moreover, sensitivity analysis of density reconstruction has been discussed for one, two and three-layered models.}
Various aspects of numerical implementation are  discussed in Sec. \ref{sec:2}. In Sec. \ref{sec:3}, we first consider toy models with simple density profiles, and then a representative density profile of the Mediterranean sea. In each case, IGWs emanating from the bottom topography impinge on the free surface.  Wavenumbers  corresponding to the  surface signature are substituted in the closed form dispersion relation to reconstruct the underlying density profile. Next we perform semi-realistic simulations of internal tides in the Strait of Gibraltar (the region where Mediterranean Sea meets the Atlantic Ocean). The first case considers real bathymetry,  real density profile, Coriolis effect, but no background velocity shear, while the second case includes the effect of velocity shear. The density stratification profile is reconstructed in both cases. 
The article is concluded in Sec. \ref{sec:4}.

\section{Governing equations and exact solutions}
\label{sec:1}
We consider an incompressible, 2D ($x-z$ plane), density stratified flow of a Boussinesq fluid on an $f$-plane. The mean (denoted by overbars) density profile varies in  the vertical ($z$) direction. 
The governing Navier-Stokes equation in this case is given by \cite{gerkema2001internal,gerkema2008introduction,kumar2018physics}:
\begin{subequations}
\begin{align}
\frac{\partial \bm{u}}{\partial t} + (\bm{u} \cdot \nabla) \bm{u} + {f\hat{z} \times \bm{u}}= 
		- \frac{1}{\rho_0} \nabla p - \frac{\rho g}{\rho_0} \hat{z} + \nu \nabla^2 \bm{u}, \label{eq1}\\
 \nabla \cdot \bm{u} = 0,  \label{eq1q}
 \\
\frac{\partial \rho}{\partial t} + \bm{u} \cdot \nabla \rho = w\frac{\rho_0}{g}N^2+
		\kappa \nabla^2 \rho + \mathcal{F}.     \label{eq2}    
\end{align}
\end{subequations}
Except $\mathcal{F}$, the variables without overbars denote the perturbation quantities.  The perturbation velocity field is denoted by  $\bm{u} \equiv (u,v,w)$, $p$ and $\rho$ respectively denote the perturbation pressure and density. 
{Since the flow consider is 2D, any variation normal to the $x-z$ plane has been neglected.
The quantity $f$ represents the Coriolis frequency, defined as $f\equiv 2 \Omega \sin \theta$, where $\Omega$ is the Earth's rotation rate (=$7.3\times10^{-5}$ s$^{-1}$) and $\theta$ is the latitude of interest.} The quantity $g$ denotes the gravitational acceleration,  $\rho_0$ represents the reference density, $\nu$ is the kinematic viscosity and $\kappa$ is the mass diffusivity. Furthermore, $N(z) \equiv \sqrt{-({g}/{\rho_0}) {d\bar{\rho}}/{dz}}$ is the Brunt-V\"ais\"al\"a (or buoyancy) frequency, which is a measure of the background stratification. The forcing function, $\mathcal{F}$, is assumed to be zero here.
In the linear regime, (\ref{eq1})--(\ref{eq2}) can be simplified into one equation by neglecting the effects of viscosity and diffusivity, and can be expressed as
\begin{equation}
\label{eq_1a}
\frac{\partial^2}{\partial t^2} \left(\frac{\partial^2 w}{\partial x^2} 
+\frac{\partial^2 w}{\partial z^2} \right) 
+{f^2\frac{\partial^2 w}{\partial z^2}}
+ N^2(z) \frac{\partial^2 w}{\partial x^2} = 0.
\end{equation}
We seek  a plane-wave solution and express $w$ as follows: $w = {W(z) \ee^{ \ii (kx -\omega t)}}$, where $k$ is the 
wavenumber in the $x$-direction and $\omega$ is the frequency. 
By substituting this ansatz in (\ref{eq_1a}), we get
\begin{equation}
\label{eq_1b}
\frac{d^2W}{dz^2} + k^2 \frac{N^2(z)-\omega^2}{\omega^2-{f^2}} W = 0.
\end{equation}
We assume the lower boundary (at $z=-H$) to be impenetrable, i.e. $w=0$ (implying $W=0$), while the upper boundary ($z=0$) to be a free surface. 
However, at the leading order approximation $w=0$ (implying $W=0$) at the free surface -- which is popularly known as the `rigid-lid approximation'\cite{gerkema2008introduction,wunsch2013baroclinic}.

Eq. (\ref{eq_1b}) together with the homogeneous boundary conditions, constitute a regular Sturm-Liouville boundary value problem. Its solution is formed by the superposition of a countably infinite set of eigenvalues $k_n$ and corresponding eigenfunctions $W_n$. The solution, which physically represents internal gravity waves, can be obtained in analytical form only for some  special choices of $N(z)$, e.g.\ constant or piecewise constant \cite{gerkema2008introduction}, 
otherwise Eq. (\ref{eq_1b}) has to be solved numerically. Below we provide the exact solutions for (i) a single layer of constant $N$, (ii) two layers, each having a constant $N$, and (iii) three layers, each having a constant $N$.
\subsection{Exact solutions}
\subsubsection{One layer}
We consider a  mean density profile $\bar{\rho}(z)$ that varies linearly with $z$, giving a constant $N$. In this situation, Eq. (\ref{eq_1b}), along with the homogeneous Dirichlet boundary conditions, can be solved exactly, yielding
\begin{equation}
\label{eq_110}
W =  \sum_{n=1}^{\infty} W_n = \sum_{n=1}^{\infty} C_n \sin \left(m_n z \right),
\end{equation}
where $m_n=n\pi/H$ is the vertical wavenumber of the $n$-th mode, and  $C_n \in \mathbb{R}$. 
The dispersion relation is given by 
\begin{equation}
\label{eq_111}
{m_n = \pm k_n\sqrt{\frac{N^2 - \omega_0^2}{\omega_0^2-f^2}}},
\end{equation}
where $k_n$ is the horizontal wavenumber of the $n$-th mode. In the above equation, we have fixed the value of $\omega$ as $\omega_0$, which we take as the tidal frequency. This is because our interest here is to obtain internal tides, that is, internal waves oscillating at tidal frequencies (IGWs). Eqs. (\ref{eq_110}) and/or (\ref{eq_111}) appear in classic texts \cite{turner1979buoyancy,gerkema2008introduction,sutherland2010internal}.
\subsubsection{Two layers}
In this case we consider a 
two-layered density stratified flow. Density in each layer varies linearly with $z$ (i.e., $N$ is constant in each layer) as follows:
\begin{eqnarray}
 \label{eq_2n}
  N = 
  \begin{cases}  
  N_1 & \   -h < z < 0, \\  
  N_2 & \   -H < z < -h. 
  \end{cases}
\end{eqnarray} 
 We note here that the two-layered density stratification is such that $\bar{\rho}(z)$ is still continuous, implying there are no \emph{interfacial} gravity waves at the pycnocline $z=-h$. In such a system,  $W_n$ can be written as \cite{gerkema2008introduction}:
 \begin{eqnarray}
 \label{eq33}
  W_n = 
  \begin{cases}  
  C_{n,1} \sin[m_{n,1} z]  & -h < z < 0, \\ \nonumber
  C_{n,2} \sin[m_{n,2} (z+H)]  & -H < z < -h. 
  \end{cases}
\end{eqnarray} 
By demanding the continuity of $W_n$ and $dW_n/dz$ at $z=-h$, we arrive at the dispersion relation 
\begin{eqnarray}
\label{eq_2nd}
m_{n,2} \sin [m_{n,1} h] \cos [m_{n,2} (H-h)] + 
m_{n,1} \cos [m_{n,1} h] \sin [m_{n,2} (H-h)] = 0,
\end{eqnarray}
where {$m_{n,i}=k_n\sqrt{N_i^2-\omega_0^2}/\sqrt{\omega_0^2-f^2}$}; $i=1,2$. 

\subsubsection{Three layers}
Next we consider a three-layered density stratified flow, with $N=\mathrm{constant}$ in each layer:
\begin{eqnarray}
 \label{eq_3n}
  N = 
  \begin{cases}  
  N_1 & \   -h_1 < z < 0, \\ 
  N_2 & \   -h_2 < z < -h_1, \\  
  N_3 & \   -H < z < -h_2. 
  \end{cases}
\end{eqnarray}
Again we note that the $\bar{\rho}(z)$ is continuous. As already mentioned, the oceans can be broadly divided into three regions of different density stratifications, hence the three-layered model can crudely represent the ocean's mean density profile. Thus $N_1$, $N_2$ and $N_3$ respectively denote the stratifications of the top, middle (pycnocline)  and bottom layers. In this three-layered system, $W_n$ can be expressed as
\begin{eqnarray}
 \label{eq3}
  W_n = 
  \begin{cases}  
  C_{n,1} \sin[m_{n,1} z]  & -h_1 < z < 0, \\ \nonumber
  C_{n,2} \sin[m_{n,2} (z+h_2)] + 
  C_{n,3} \cos[m_{n,2} (z+h_2)] & 
 -h_2 < z < -h_1, \\ \nonumber
  C_{n,4} \sin[m_{n,3} (z+H)] 
  & -H < z < -h_2,
  \end{cases}
\end{eqnarray} 
where {$m_{n,i}=k_n\sqrt{N_i^2-\omega_0^2}/\sqrt{\omega_0^2-f^2}$}; $i=1,2,3$.
To obtain the four unknown coefficients, we  demand the continuity of $W_n$ and $dW_n/dz$ at the two interfaces $z=-h_1$ and $z=-h_2$, which finally yields the dispersion relation 
\begin{eqnarray}
\label{eq5}
m_{n,2} m_{n,3}   \sin[m_{n,1} h_1] \cos[m_{n,2}(h_1-h_2)] \cos[m_{n,3}(H-h_2)]  \nonumber \\
- m_{n,1} m_{n,3} \cos[m_{n,1} h_1] \sin[m_{n,2}(h_1-h_2)] \cos[m_{n,3}(H-h_2)]  \nonumber \\
+ m_{n,1} m_{n,2} \cos[m_{n,1} h_1] \cos[m_{n,2}(h_1-h_2)] \sin[m_{n,3}(H-h_2)]  \nonumber \\
+ m_{n,2}^2       \sin[m_{n,1} h_1] \sin[m_{n,2}(h_1-h_2)] \sin[m_{n,3}(H-h_2)]   = 0.
\end{eqnarray}

\subsection{Uniqueness of the wavenumbers}
Our objective is to reconstruct the ocean's density stratification $N$ by only analyzing the free surface data.  Due to the tidal forcing,  internal wave beams radiate from the bottom topography and the low mode internal waves impinge on the free surface. Spectral analysis of the free surface would reveal the wavenumbers $k_n$, which, on being substituted into the $m$-layered dispersion relation (for example, the $3$-layered dispersion relation is given by Eq. (\ref{eq5})) would reconstruct $N$ as an $m$-layered profile. However, a fundamental mathematical question in this regard is -- \emph{is this reconstruction unique}? In other words, can two different stratification profiles have exactly the same set of wavenumbers?  

\subsubsection{Uniqueness of isolated eigenvalues of Sturm-Liouville boundary value problems}

Let us consider a regular, two point, Sturm-Liouville boundary value problem
\begin{align*}
-\frac{d}{dz} \bigg[p(z) \frac{dY}{dz}\bigg] + r(z)Y  = k w(z) Y; & \qquad z \in [a,b], \nonumber \\
\text{with boundary conditions}\,\, C_1Y(a) + C_2\frac{dY(a)}{dz} = 0 &\,\, \text{and} \,\,C_3Y(b) + C_4\frac{dY(b)}{dz} = 0.
\end{align*}
The above problem has a countable infinite number of eigenvalues $k_n$ and corresponding eigenfunctions $Y_n$. 
\citet{zettl2010sturm} (see p53 and Theorem 3.5.1, p54) shows that if two eigenvalue problems are `close' to each other, the isolated eigenvalues $k_n$ are also close. In other words, a small change in the Sturm-Liouville problem also leads to a small change in the  eigenvalues and the eigenfunctions. 
Thus the set of wavenumbers given by Eq. (\ref{eq_1b}) for a given $N$ profile (and depth $H$) must be unique. 

{In practice, however, only the first few-modes can be inferred with reasonable accuracy from the satellite altimetry measurements\cite{zhao2012mapping}. A fairly accurate density reconstruction is still possible with a finite number of modes. The accuracy of reconstruction increases as we collect more modes (see Fig. \ref{fig:err} and Sec. \ref{realistic} for detailed discussion). }

\subsubsection{Reflection symmetry of the density profile: A consequence for eigenvalue uniqueness}
\label{symmetry}
Here we show that under symmetry transform, the Sturm-Liouville problem may remain unchanged, and hence provides the same set of eigenvalues. We focus on the specific Sturm-Liouville problem concerning IGWs in Eq. (\ref{eq_1b}), and apply the reflection symmetry transform: $z \rightarrow -z-H$. This yields the exactly same Sturm-Liouville problem as Eq. (\ref{eq_1b}):
\begin{equation}
\label{eq_ref_sem}
\frac{d^2}{dz_*^2}W(z_*) + k^2 \frac{N^2(z_*)-\omega^2}{\omega^2-{f^2}} W(z_*) = 0,
\end{equation}
with boundary conditions $W(z_*)=0$ at $z_*=0$ and $z_*=-H$, where $z_*=-z-H$. Note that  $N^2(z_*)=N^2(-z-H)$, implying that $N^2(z)$ and $N^2(-z-H)$ yield the exact same set of eigenvalues $k_n$.
Two different stratification profiles (one being a reflection symmetry of  the other) yielding the same set of eigenvalues can have consequences in our 
inverse reconstruction technique. The pycnocline in the `false profile' would appear near the ocean bottom. This profile being physically unfeasible (i.e. extraneous solution) should be  disregarded.

\subsection{Sensitivity analysis}
\label{appD}

\subsubsection{One-layer model}\label{appD1}
For one-layer model, the dispersion relation can be written as
\begin{equation}
    \mathcal{D} (k,N) = \frac{n \pi}{H} - k \frac{\sqrt{N^2-\omega_0^2}}{\sqrt{\omega_0^2-f^2}} = 0.
\end{equation} 
Suppose, due to measurement inaccuracies, instead of obtaining the exact wavenumber $k$, we obtain some other value $k_1$ where $k_1=k+\delta k$. Because of this measurement error, we do not get the exact value $N$ but $N_1=N+\delta N$.
Therefore, we impose
\begin{equation}
    \mathcal{D} (k+\delta k, N+\delta N) = 0.
\end{equation} 
For small values of $\delta k$ and $\delta N$, we can do Taylor series expansion of the above equation as 
\begin{equation}
    \cancelto{0}{\mathcal{D}(k,N)} 
    + \left.\frac{\partial \mathcal{D}}{\partial k}\right\vert_{(k,N)}
    \delta k + \left.\frac{\partial \mathcal{D}}{\partial N}\right\vert_{(k,N)}
    \delta N = 0.
\end{equation} 
After some algebraic manipulations, it can be shown that
\begin{equation}
    \frac{\delta k}{k} = \left( \frac{N^2}{N^2-\omega_0^2} \right) \frac{\delta N}{N}.
\end{equation} 

In a real oceanic environment, the typical range of $N$ is $10^{-1}-10^{-3}$ s$^{-1}$
and $\omega_0$ is $1.4\times10^{-4}$ s$^{-1}$. In this setting, we can approximate ($N^2-\omega_0^2$) as $N^2$. Therefore, $\delta N/N \approx \delta k/k$, which means that the estimation error of $N$ grows linearly with the measurement error of $k$. 

\subsubsection{Two-layered model}\label{appD2}
A similar type of mathematical argument can be built up from the simple one-layer case. The number of unknowns  needed to construct the density profile in the two-layered case is $4$ ($k$, $N_1$, $N_2$ and $h$), and hence the problem is   not as the straightforward as the one-layer case. We  take an analytical approach and therefore consider a series of assumptions.
For simplicity, let us assume that the depth $h$ of the upper layer  (having stratification $N_1$) is known. The dispersion relation can be written as
\begin{equation}
    \mathcal{D} (k+\delta k, N_1+\delta N_1, N_2+\delta N_2) = 0,
\end{equation} 
where $\delta k$ is the error in the measurement of $k$; $\delta N_1$ and $\delta N_2$
are the respective  estimation errors for $N_1$ and $N_2$.
For small values of $\delta k$, $\delta N_1$ and $\delta N_2$, we can perform Taylor series expansion of the above equation as
\begin{equation}
    \cancelto{0}{\mathcal{D}(k,N_1, N_2)} 
    + \left.\frac{\partial \mathcal{D}}{\partial k}\right\vert_{(k,N_1,N_2)}
    \delta k + \left.\frac{\partial \mathcal{D}}{\partial N_1}\right\vert_{(k,N_1,N_2)} \delta N_1 + \left.\frac{\partial \mathcal{D}}{\partial N_2}\right\vert_{(k,N_1,N_2)} \delta N_2 = 0.
\end{equation} 
Further, let us  assume that the errors of  $N_1$ and $N_2$ are equal i.e., $\delta N_1=\delta N_2=\delta N$. Therefore, the above equation can be written as
\begin{equation}
    \left.\frac{\partial \mathcal{D}}{\partial k}\right\vert_{(k,N_1,N_2)}
    \delta k = - \left.\frac{\partial \mathcal{D}}{\partial N_1}\right\vert_{(k,N_1,N_2)} \delta N - \left.\frac{\partial \mathcal{D}}{\partial N_2}\right\vert_{(k,N_1,N_2)} \delta N.
\end{equation} 
After some algebraic manipulations and assuming that $N_1^2-\omega_0^2 \approx N_1^2$, $N_2^2-\omega_0^2 \approx N_2^2$, it can be shown that
\begin{equation}
    \frac{\delta k}{k} \approx \frac{\sin{[m_1 h + m_2 (H-h)]}}
    {N_1 \cos{[m_1 h]} \sin{[m_2 (H- h)]} + N_2 \sin{[m_1 h]} \cos{[m_2 (H-h)]}} 
    \delta N,
\end{equation}
where $m_i={k N_i}/{\sqrt{\omega_0^2-f^2}}$ for $i=1,2$.
From the above equation, we can recover the sensitivity of the one-layer model
by substituting $N_1=N_2$ (therefore $m_1=m_2$). 

\subsubsection{Three-layered model}\label{appD3}
{For simplicity, we assume that the depth $h_1$ of the upper layer  (having stratification $N_1$) and depth $h_2$ of the middle layer (having stratification $N_2$) are known.}
Therefore, the dispersion relation for the three-layered model can be written as 
\begin{equation}
    \mathcal{D} (k+\delta k, N_1+\delta N_1, N_2+\delta N_2, N_3+\delta N_3) = 0,
\end{equation} 
where $\delta k$ is the error in the measurement of $k$; $\delta N_1$, $\delta N_2$ and $\delta N_3$ are the respective  estimation errors for $N_1$, $N_2$ and $N_3$.
For small values of $\delta k$, $\delta N_1$, $\delta N_2$ and $\delta N_2$, we can perform Taylor series expansion of the above equation as
\begin{equation}
    \cancelto{0}{\mathcal{D}(k,N_1, N_2, N_3)} 
    +\left.\frac{\partial \mathcal{D}}{\partial k}\right\vert_{(k,N_1,N_2, N_3)} \delta k 
    + \sum_{i=1}^{3} \left.\frac{\partial \mathcal{D}}{\partial N_i}\right\vert_{(k,N_1,N_2,N_3)} \delta N_i  = 0.
\end{equation} 
Further, let us  assume that the errors of  $N_1$, $N_2$ and $N_3$ are equal i.e., $\delta N_1=\delta N_2=\delta N_3=\delta N$. Therefore, the above equation can be written as
\begin{equation}
    \left.\frac{\partial \mathcal{D}}{\partial k}\right\vert_{(k,N_1,N_2, N_3)} \delta k = 
    - \delta N \sum_{i=1}^{3} \left.\frac{\partial \mathcal{D}}{\partial N_i}\right\vert_{(k,N_1,N_2,N_3)}.
\end{equation} 
After some algebraic manipulations and assuming that $N_1^2-\omega_0^2 \approx N_1^2$, $N_2^2-\omega_0^2 \approx N_2^2$, $N_3^2-\omega_0^2 \approx N_3^2 \approx N_1^2$, it follows that
\begin{align}
 & N_1 N_2 \cos[m_2h] \cos[m_1(H+h)] \frac{\delta k}{k}
 \approx 
\nonumber \\
 &\frac{\delta N}{N_1} \Big[N_1 N_2 \cos[m_1(H+h)] \cos[m_2h]+m_1 N_2 \frac{h}{H} \cos[m_1(H+h)+m_2h] 
\nonumber \\
 & -N_1^2 \cos[m_1h_1+m_2h] \left(\frac{h_1}{H}\cos[m_1(H-h_2)]-\frac{h_2}{H} \cos[m_1(H-h_1)]\right)
\nonumber \\
 & +N_2^2 \sin[m_1h_1+m_2h] \left(\frac{h_1}{H}\sin[m_1(H-h_2)] -\frac{h_2}{H}\sin[m_1(H-h_1)]\right)
\nonumber \\
 & +\frac{N_1+N_2}{kH} \sqrt{\omega_0^2-f^2} \sin[m_1(H+h)] \cos[m_2h]\Big],
\end{align}
where $h=h_1-h_2$ and $m_i={k N_i}/{\sqrt{\omega_0^2-f^2}}$ for $i=1,2$.
For a thin pycnocline ($h\approx0$, implies $h_1\approx h_2$), the above equation can be simplified as
\begin{equation}
    \frac{\delta k}{k} \approx
    \frac{\delta N}{N_1} \left( 1 + \left[\frac{1}{m_1 H}+\frac{1}{m_2 H}\right]\tan[m_1H] \right). 
\end{equation}

\section{Numerical implementation}
\label{sec:2}
In order to simulate the internal tides, we numerically solve equation set Eqs. (\ref{eq1})--(\ref{eq2}). Following Gerkema \cite{gerkema2001internal}, we  consider the barotropic tidal forcing term
\begin{equation}
\mathcal{F} = z N^2(z) \frac{Q_0 \sin(\omega_0 t)}{{h(x)}^2} \frac{dh}{dx},
\end{equation}
where $Q_0$ is the barotropic flux,  $h(x)$ is the local water depth and $\omega_0$ is the tidal frequency.  
Th bottom topography has been incorporated; furthermore, the Cartesian coordinate system, $x-z$, has been transformed  into a terrain-following coordinate system, $x-\zeta$ with $\zeta \equiv -z/h(x)$. 
Therefore, in the terrain-following coordinate system, the undisturbed free surface  is denoted by $\zeta=0$, and the bottom surface lies at $\zeta=-1$. 
Following Dimas and Triantafylluon \cite{dimas1994nonlinear}, we use a spectral spatial discretization with Chebyshev polynomial in the vertical direction and Fourier modes in the streamwise direction, the latter has been assumed to be periodic. Eqs. (\ref{eq1})--(\ref{eq2}) have been solved using an open-source pseudo-spectral code -- Dedalus \cite{burns2019dedalus}. Fourth-order Runge-Kutta method has been used for the time-marching.

The correspondence between the prognostic variables in the two coordinate systems are: $\tilde{\bm{u}}(x,\zeta,t) = \bm{u}(x,z,t);\, \tilde{p}(x,\zeta,t) = p(x,z,t);\,\tilde{\rho}(x,\zeta,t) = \rho(x,z,t)$.
At the free surface $\zeta=\eta(x,t)$, the kinematic and dynamic boundary conditions are respectively given by
\renewcommand{\theequation}{\arabic{section}.\arabic{equation}a,b}
\begin{equation}
\tilde{w} = \frac{\partial \eta}{\partial t} + \tilde{u} \frac{\partial \eta}{\partial x} \qquad;\qquad
\tilde{p} = 0,
\end{equation}
\renewcommand{\theequation}{\arabic{section}.\arabic{equation}}
where $\eta$ is the free surface elevation. At the free surface, $\tilde{u}$ satisfies the stress free boundary condition. At the bottom surface $\zeta=-1$,  $\tilde{u}$ and $\tilde{w}$ respectively satisfies the no-slip and the no-penetration boundary conditions. 
The insulating boundary conditions applied both at the top and the bottom have been used for the density. Furthermore, we have taken both viscosity and diffusivity into account, $\nu$ and $\kappa$ are respectively set to $10^{-6}$ m$^2$s$^{-1}$ and $10^{-7}$ m$^2$s$^{-1}$.

For numerical simulations, we have first considered a toy model with a \textit{Gaussian} bottom topography. The density profile of the model has been varied from single to three-layered (given in Sec. \ref{sec:3.1}--\ref{sec:3.3}).   
The model has a depth of $H=1$ m, a horizontal extent of $40$ m, and has been forced with a barotropic flux $Q_0=10^{-3}$ m$^2$ s$^{-1}$ and frequency  $\omega_0=0.05$ s$^{-1}$. Since the topography radiates IGWs,  sponge layers have been used to absorb the incoming IGWs both at the east and the west boundaries of the domain. We have used $256$ Chebyshev points in the $z$-direction and $1024$ Fourier-modes along the $x$-direction. We have simulated $8$ tidal periods with a time-step of $0.1$ s.

Next we have considered a slightly more realistic scenario in Sec. \ref{sec:3.4} and studied the IGWs generation in  the tidally active part of the Mediterranean sea. The stably stratified, time-averaged and smoothed density profile has been taken at $36.6^\circ$ N  latitude and $0.2^\circ$ W longitude. To simulate this we have used a domain of ($L_x \times L_z$)  = ($50 \times 1$) km with sponge layers of $10$ km on both eastern and western boundaries. The flow is forced using semi-diurnal tides of frequency  $1.4\times10^{-4}$ s$^{-1}$  and with $Q_0=10^{-3}$ m$^2$s$^{-1}$ over a Gaussian mountain. The spatial and temporal discretization, as well as the total time are same as that in Sec. \ref{sec:3.1}--\ref{sec:3.3}.

\section{Results}
\label{sec:3}
\subsection{Toy model}
\subsubsection{One-layer}
\label{sec:3.1}
We first simulate a single-layered flow with $N=0.1$ s$^{-1}$ and {$f=0$ s$^{-1}$}. The space-time Fourier Transform (STFT) of the surface elevation field $\eta$ yields the wavenumbers $k_n$ corresponding to the tidal frequency $\omega_0=0.05$ s$^{-1}$.  Figure \ref{fig:one}(a) shows STFT of the free surface elevation; the first vertical-mode ($n=1$) corresponds to $k_1=1.813$ m$^{-1}$. By substituting $\omega_0$ and $k_1$ in (\ref{eq_111}), we straightforwardly estimate $N$. Since the density at the surface is known, the mean density profile $\bar{\rho}(z)$ can be directly reconstructed; see figure \ref{fig:one}(b). Moreover, this result also serves as a validation of the numerical code. 
{Sensitivity analysis (see, Sec. \ref{appD1}) shows that error in estimation of $N$ approximately grows with measurement error of $k_n$. For this particular case, $\delta N/N=0.75 (\delta k/k)$. }

\subsubsection{Two-layered}
\label{sec:3.2}

\begin{figure*}
\centering
\includegraphics[width=1.0\linewidth]{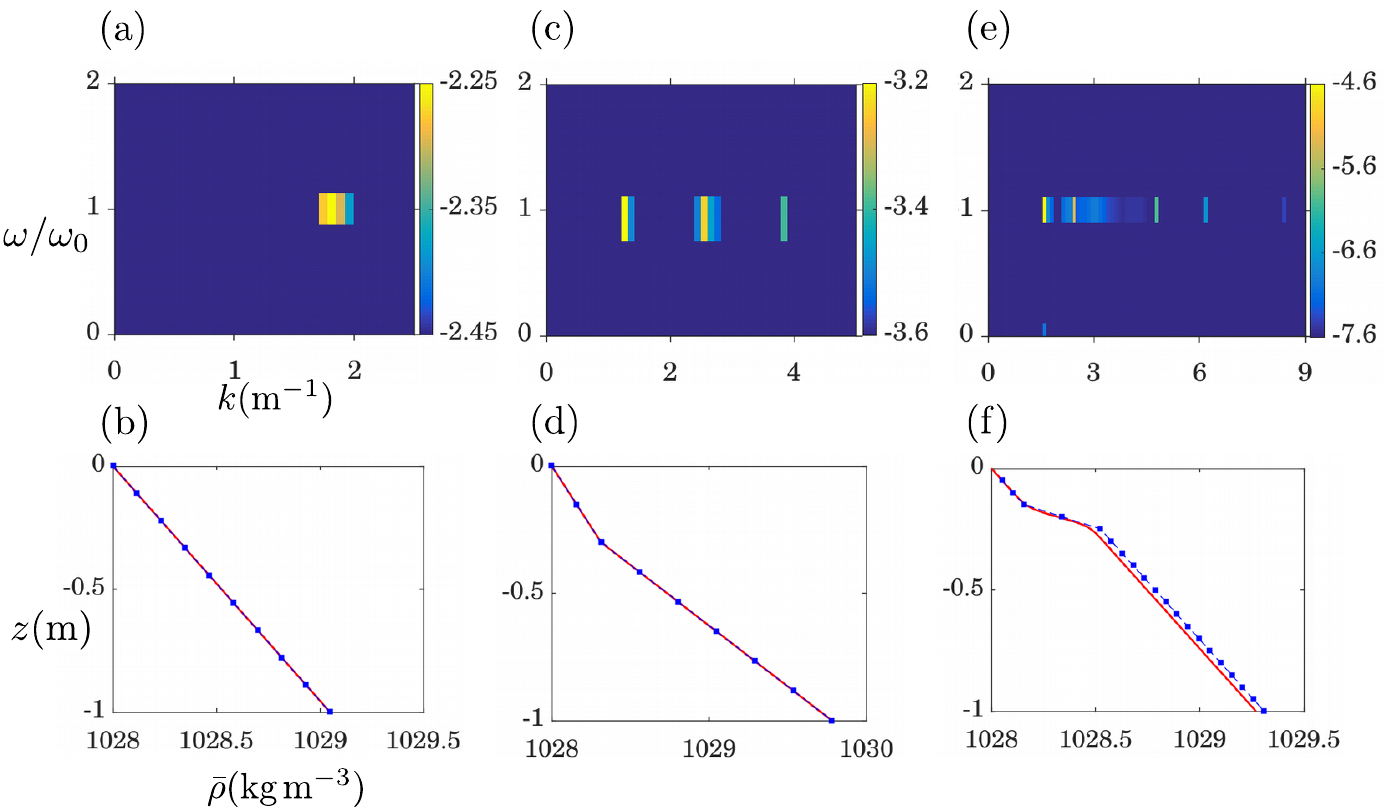}
\caption{Top panel shows STFT of the free surface displacement (contours represent amplitudes in log-scale), while bottom panel shows comparison between the actual (red line) and the estimated (blue dashed line with  markers) mean  density profiles. (a)-(b) one-layer, (c)-(d) two-layer and
(e)-(f) three-layer.  }
\label{fig:one}
\end{figure*}

Here we consider a squared buoyancy frequency
\begin{equation}
\label{eq_1c}
N^2 = 2\times 10^{-2} - 10^{-2} \frac{1}{1+(z-0.7)^{256}}, 
\end{equation}
which approximates Eq. (\ref{eq_2n}) with $N_1=0.1$ s$^{-1}$, $N_2=0.14$ s$^{-1}$ and $h=0.3$ m. Similar to the one-layer case in Sec. \ref{sec:3.1}, tidal forcing leads to the IGWs radiation, whose imprint is detectable at the free surface. { In this case, we also fix $f=0$ s$^{-1}$.} Fig. \ref{fig:one}(c) shows the  STFT of the free surface elevation; corresponding to the tidal frequency, the first three vertical modes respectively peak at $k_1=1.413$ m$^{-1}$, $k_2=2.549$ m$^{-1}$ and $k_3=3.836$ m$^{-1}$. Our objective is to reconstruct Eq. (\ref{eq_1c}) using Eq. (\ref{eq_2nd}), which means that the three unknowns, $N_1$, $N_2$ and $h$ have to be evaluated. We substitute the obtained values of $k_1$, $k_2$ and $k_3$ along with $\omega_0$ in Eq. (\ref{eq_2nd}),  leading to a system of three equations and three unknowns, which is then solved numerically.  {The estimated values of $N_1$, $N_2$ and $h$ are respectively 0.095 s$^{-1}$, 0.132 s$^{-1}$ and $0.29$ m and with these values the density profile has been reconstructed.} 
The mean density profile (solid line) along with the reconstructed version (blue dashed line with markers) are shown in Fig. \ref{fig:one}(d). 
{The sensitivity analysis (see, Sec. \ref{appD2}) suggests that
$\delta N/N_{avg}=1.58 (\delta k/k)$, where $N_{avg}=(N_1+N_2)/2$. 
}

\subsubsection{Three-layered}
\label{sec:3.3}

\begin{figure}
\centering
\includegraphics[width=0.9\linewidth]{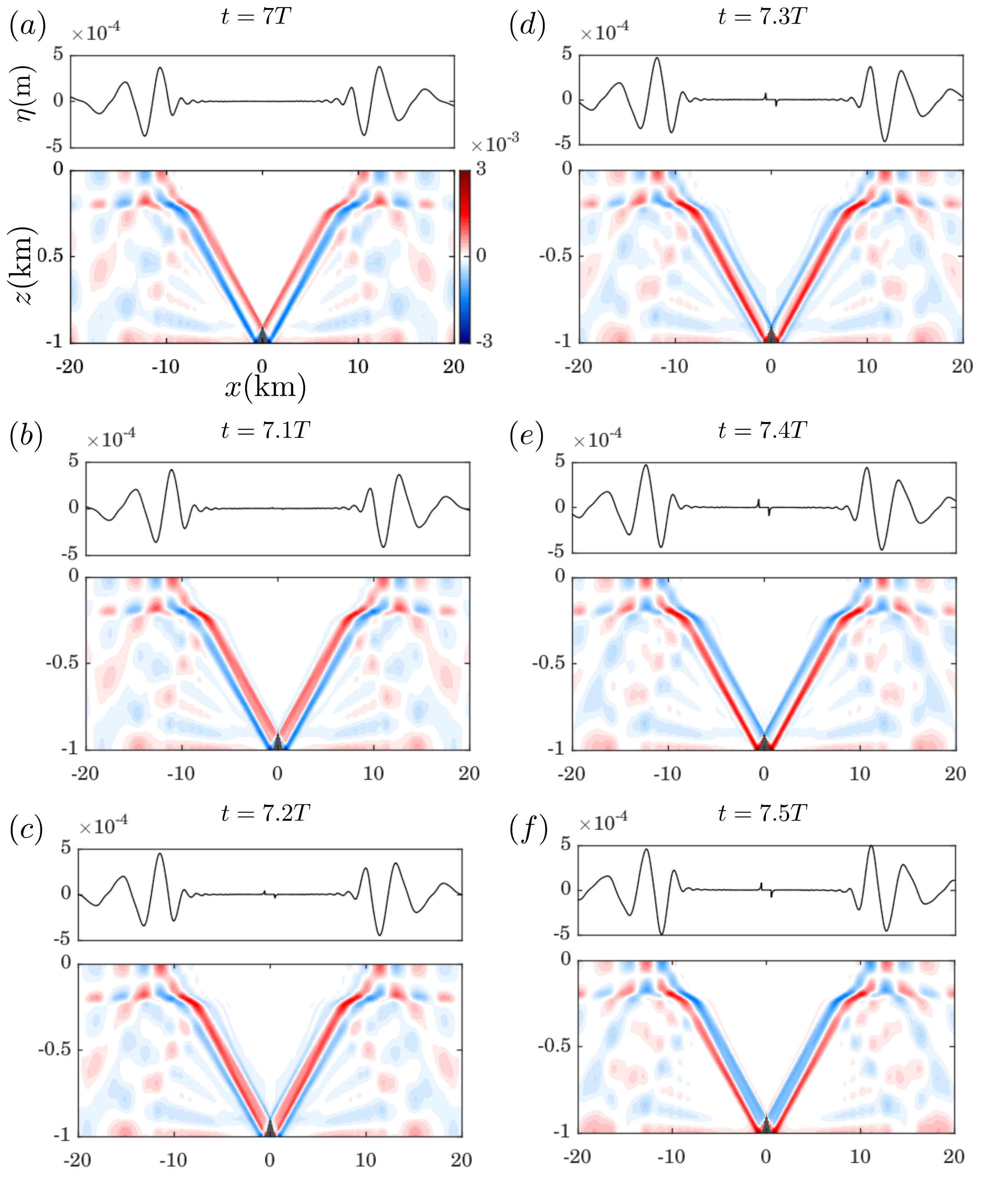}
\caption{The IGWs radiating from a Gaussian submarine mountain impinges on the free surface. Density profile, representative of the Mediterranean sea, has been considered. Snapshots of  the free surface displacement ($\eta$, in m) and the corresponding horizontal baroclinic velocity field ($u$, in ms$^{-1}$) are shown. Time corresponding to each snapshot appears at the top of each sub-figure. The semi-diurnal tidal period, $T = {2\pi}/\omega \approx 12.46$ hours.}
\label{fig:two}
\end{figure}

\begin{figure}
\centering
\includegraphics[width=0.9\linewidth]{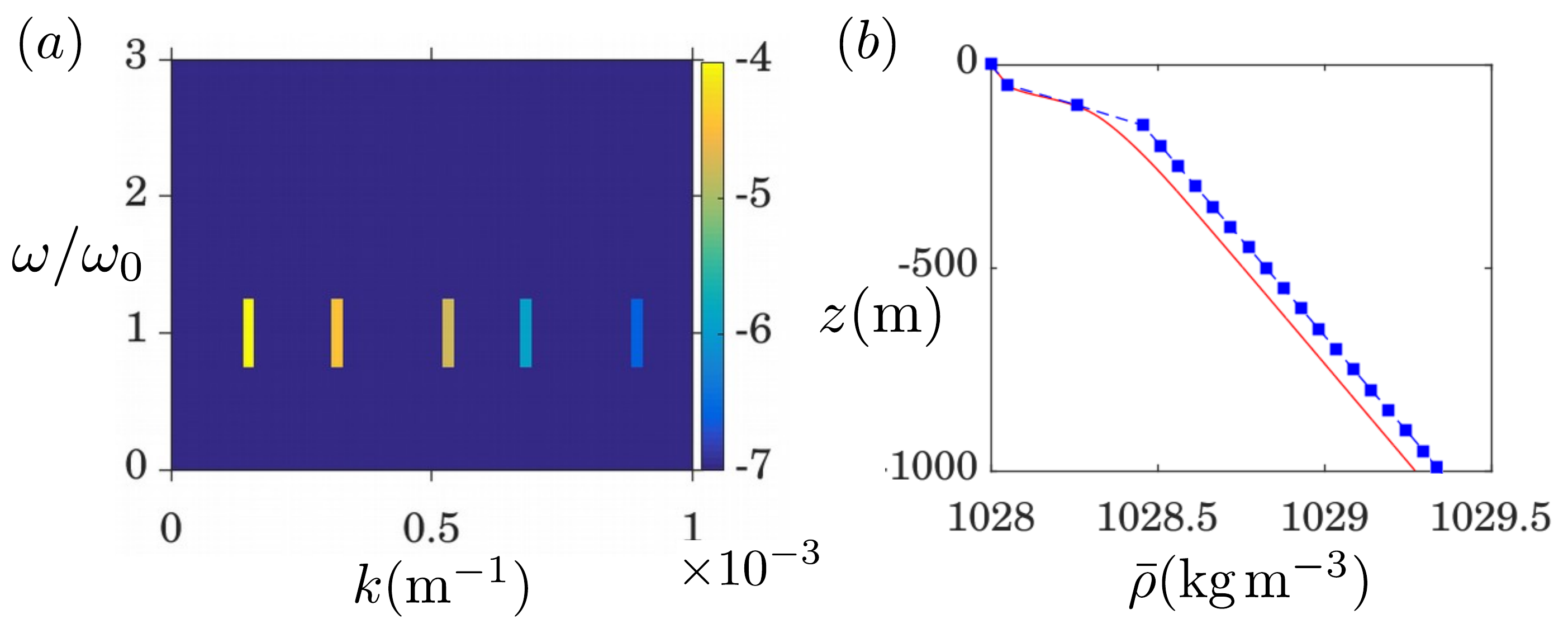}
\caption{ Density reconstruction for the Mediterranean sea profile. (a)  STFT of free surface displacement (contours represent amplitudes in log-scale), and (b)  actual (red line) and the estimated (blue dashed line with  markers) mean  density profile.}
\label{fig:three}
\end{figure}

We follow the same strategy as that outlined in Sec. \ref{sec:3.1} and Sec. \ref{sec:3.2}. In this case, we use the following $N^2$ profile:
\begin{equation}
\label{eq_6}
N^2 = 0.01 + 0.0625 \exp(-1000\times(-z+0.2)^2). 
\end{equation}
The above profile approximates the three-layered configuration of Eq. (\ref{eq_3n}), with $N_1=0.1$ s$^{-1}$, $N_2=0.25$ s$^{-1}$, $N_3=0.1$ s$^{-1}$, {$f=0$ s$^{-1}$}, $h_1=0.12$ m and $h_2=0.28$ m. The goal is to find $N_1$, $N_2$, $N_3$, $h_1$ and $h_2$, and therefore we construct five-equations from Eq. (\ref{eq5}) for different value of $n$. 
Fig. \ref{fig:one}(e) reveals that $k_1=1.396$ m$^{-1}$, $k_2=2.448$ m$^{-1}$, $k_3=4.789$ m$^{-1}$, $k_4=6.173$ m$^{-1}$ and $k_5=8.408$ 
m$^{-1}$. The system of five equations and five unknowns resulting from Eq. (\ref{eq5}) are then solved numerically. {The estimated values of $N_1$, $N_2$, $N_3$, $h_1$ and $h_2$ are respectively 0.093 s$^{-1}$, 0.246 s$^{-1}$, 
0.096 s$^{-1}$, $0.1$ m and $0.32$ m. The density profile has been reconstructed
with the estimated parameters.} The mean density profile (solid line) along with the reconstructed version (blue dashed line with markers) are shown in Fig. \ref{fig:one}(f).
{The sensitivity analysis (see, Sec. \ref{appD3}) suggests that $\delta N/N_{avg}=1.89 (\delta k/k)$, where $N_{avg}=(N_1+N_2+N_3)/3$. 
}

\subsection{Mediterranean sea profile}

\subsubsection{Idealistic Gaussian topography}
\label{sec:3.4}
 
As already mentioned in Sec. \ref{sec:2}, we simulate the IGWs for a case in which the mean density profile is representative of the Mediterranean sea.  We have assumed $f=0$ s$^{-1}$ for simplicity.  
Fig. \ref{fig:two} (also see supplementary Movie) shows snapshots of the free surface displacement $\eta$ along with the contours of the horizontal baroclinic velocity $u$ in the vertical plane at different time instants. Bending of the internal beams occur due to refraction from the pycnocline, furthermore, reflection from the pycnocline (which is  of moderate strength) leads to the observed beam scattering \cite{gerkema2001internal}. A point worth mentioning here is that the wavenumbers present at the free surface  (and hence, in the internal beam) do not depend on the underlying topography at the generation site. Therefore, the result should be valid for any other bottom topography that is not flat. Fig. \ref{fig:three}(a) represents the STFT of the free surface elevation, the first five vertical-modes are $k_1=1.49\times10^{-4}$ m$^{-1}$, $k_2=3.19\times10^{-4}$ m$^{-1}$, $k_3=5.32\times10^{-4}$ m$^{-1}$, $k_4=6.81\times10^{-4}$ m$^{-1}$ and $k_5=8.94\times10^{-4}$ m$^{-1}$. We approximate the Mediterranean sea profile with a three-layered model, and hence follow the same procedure mentioned in Sec. \ref{sec:3.3}. 
The estimated values of $N_1$, $N_2$, $N_3$, $h_1$ and $h_2$ are respectively $1.1\times10^{-3}$ s$^{-1}$, $6.5\times10^{-3}$ s$^{-1}$, $2.5\times10^{-3}$ s$^{-1}$, $50$ m and $150$ m. 
Fig. \ref{fig:three}(b) shows the  comparison between the actual density profile used in the model and the estimated density profile. The sensitivity analysis (see, Sec. \ref{appD3}) suggests that $\delta N/N_{avg}=2.81 (\delta k/k)$, where $N_{avg}=(N_1+N_2+N_3)/3$. 

To estimate the error, we have used  normalised root-mean-squared error (NRMSE), defined as $$NRMSE=\frac{\sqrt{\frac{1}{M}\sum_{i=1}^{M} (\bar{\rho}^{a}_i-\bar{\rho}^{e}_i)^2}}
{\bar{\rho}^{a}_{max}-\bar{\rho}^{a}_{min}},$$ where
$\bar{\rho}^{a}$ and $\bar{\rho}^{e}$ are respectively the actual and the estimated
density profiles, $\bar{\rho}^{a}_{max}$ and $\bar{\rho}^{a}_{min}$
are  respectively the maximum and the minimum values of the actual density profile,
and $M$ is the number of points.
We find NRMSE to be $5.4\%$,
which implies that the three-layered model (or in other words, the first five modes) estimates the density profile with reasonable accuracy. 
\subsubsection{Realistic topography without and with shear}
\label{realistic}
\begin{figure}
\centering
\includegraphics[width=0.85\linewidth]{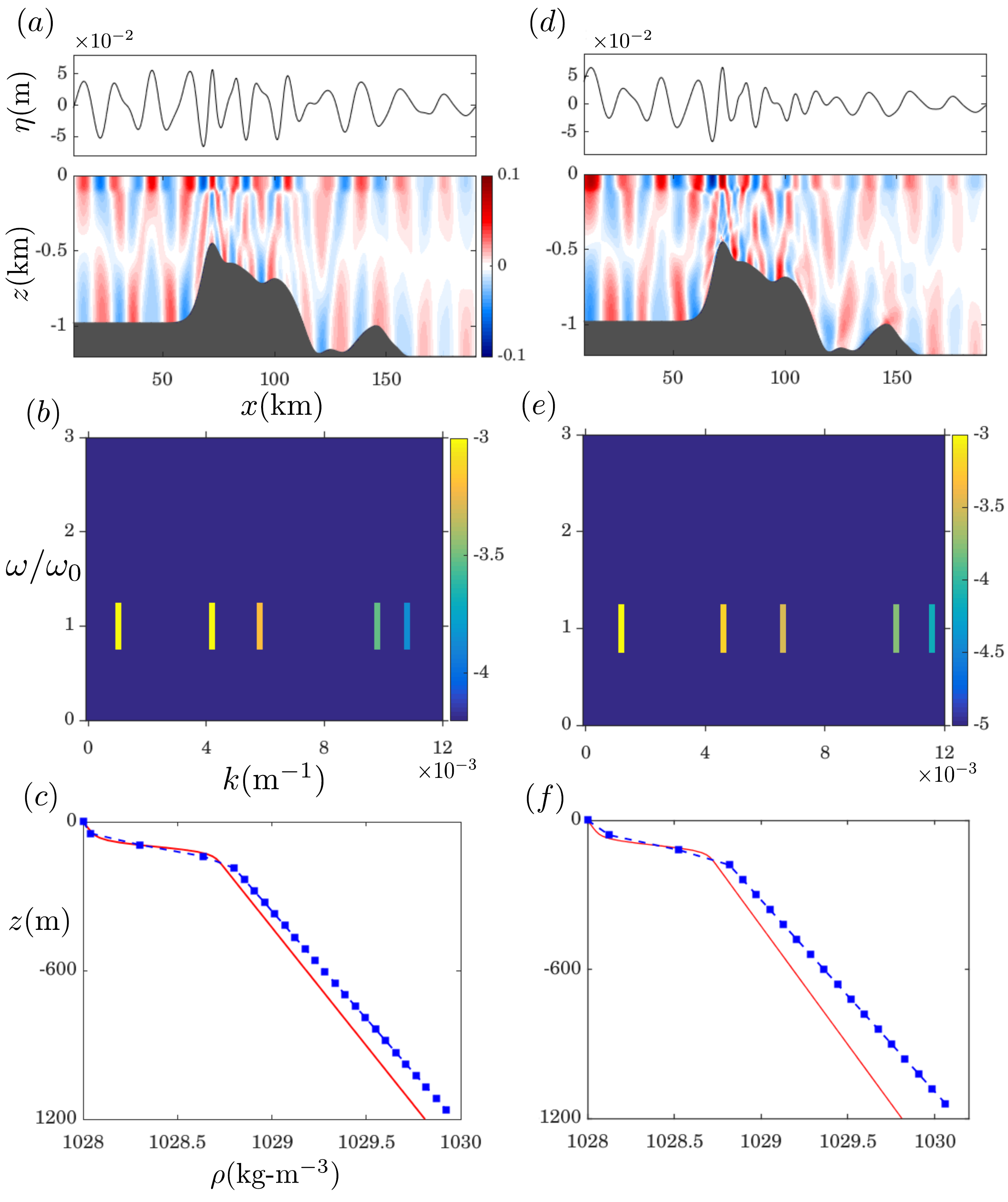}
\caption{Simulation of IGWs near the Strait of Gibraltar. The left column  ((a)--(c)) is the no-shear case, while the right column ((d)--(f)) corresponds to a steady background shear. The top panel  ((a) and (d))shows  snapshots of the free-surface displacement ($\eta$, in m) and the corresponding horizontal baroclinic velocity contours ($u$, in ms$^{-1}$). The middle panel ((b) and (e)) shows the STFT of the free-surface displacement (contours represent amplitudes in log-scale), while the bottom panel ((c) and (f))  shows the actual (red line) and the estimated (blue dashed line with  markers) density profile.}
\label{fig:shear}
\end{figure}

{It has been observed that the Strait of Gibraltar is a \emph{choke-point} in the exchange of water between the Atlantic Ocean and the Mediterranean Sea \cite{send2001intensive}. 
Therefore, it is possible that there might be a dynamical role of the shear flow on the properties of internal tides (i.e.\ the dispersion relation might be affected). 
Based on  observational evidences, we pursue this issue by 
performing two numerical experiments -- one without background shear and another with  steady background shear. While in the Strait of Gibraltar, the shear is primarily caused by the exchange flow, in a different situation it can be caused by a strong wind forcing; hence the imposition  of the background shear can be viewed as a general case. 
In this regard, we have solved the $2$D Navier Stokes equations with Boussinesq
approximation on an $f$-plane with  realistic ocean topography. 
The  bathymetry considered here is  located at $30\degree$N latitude
and along the Strait of Gibraltar. The bathymetry data has been taken from  the General Bathymetric Chart of the Oceans’ (GEBCO) gridded bathymetric datasets \cite{weatherall2015new} and interpolated to the numerical grid resolutions. For the numerical simulation purposes, we have used MITgcm, an open-source code \cite{marshall1997finite}. The horizontal length of the computational domain is $200$ km with $2000$ uniform grid points. In vertical direction we have used $200$ non-uniform grid points and nearly $20$ grid points has been concentrated near to the bottom boundary to fairly resolves the boundary layer. Both kinematic viscosity and mass diffusivity have been set to $10^{-6}$ m$^{2}$ s$^{-1}$, and sub-grid modeling hasn't been used. No-slip and no-penetration velocity boundary conditions are applied at the bottom of the domain and no-flux boundary conditions are applied to the density field at the ocean surface and at the bottom. The numerical model incorporates implicit free surface with partial-step topography formulation \cite{adcroft1997representation}.
For generating the internal tides, the model has been forced with semi-diurnal
barotropic tide with  a frequency of $1.4\times10^{-4}$ s$^{-1}$. The model has been integrated up to $7$ tidal periods with a time-step of $20$ seconds.
The shear flow profile used in this study is taken from the observation data reported in Send and Baschek \cite{send2001intensive} (see, solid black line figure 7(a)   of their paper). The steady shear flow has been  calculated from the total shear flow by removing the constant barotropic tide, which is $0.05$ ms$^{-1}$ (see \citet{izquierdo2019role}). 
The maximum and minimum value of the shear flow is found to be $0.35$ and $-0.15$  
ms$^{-1}$ respectively. However, stratified shear flows may undergo  instability 
if the value of the local Richardson number ($Ri\equiv N^2/(dU/dz)^2$ where $U(z)$ is the imposed shear velocity) is lesser than $1/4$ (see \citet{draz1982}). In our  numerical simulation, $\min(N)=5\times10^{-3}$ s$^{-1}$ and $\max(dU/dz)=5\times10^{-3}$  s$^{-1}$, hence $\min (Ri)=1$, making the background state linearly stable.
Both Figs. \ref{fig:shear}(a) and \ref{fig:shear}(d) show  snapshots of the surface undulation created by the IGWs and the perturbed baroclinic $u$-velocity contours; the first figure is without and the second one is with the shear. While the coherent internal beam is observed in the $u$-velocity contours of Fig. \ref{fig:shear}(a) (just like Fig. \ref{fig:two}),  slight loss of coherence due to the moderate amount of shear is observed in Fig. \ref{fig:shear}(d). The effect of shear becomes clearer on comparing the STFT of the two cases. Figs \ref{fig:shear}(b) and \ref{fig:shear}(e) respectively show the wavenumbers of the first $5$ modes for without and with shear cases, and their numerical values have been given in Table \ref{tab:k_n}.
A simple comparison of these values suggests that the moderate shear has negligible effect on the wavenumber of  mode-$1$.   However, as expected, the shear has some effect (albeit small) on the higher modes; higher the mode number, higher is the change. Moreover, such changes would also increase with increasing shear, and there is a possibility to loose the coherent beam structure when the shear is very strong. Table \ref{tab:k_n} reveals that the wavelength of the first $5$ modes range from $\sim 100$ km to $\sim 10$ km; it is difficult for shear, which works at $10$--$100$ m scales, to have a  strong effect on these modes. Since our density reconstruction procedure is based on `reading' wavenumbers from the free surface and using these wavenumbers as an input in Eq. (\ref{eq5}) (which has been derived for the no-shear condition), changes in the wavenumbers would reflect in the estimation of the density profile. 
{Table \ref{tab:N_n} shows the estimated values of required parameters  to reconstruct the density field using the theory of three-layered model.}
Figs. \ref{fig:shear}(c) and \ref{fig:shear}(f) respectively show the estimation of density profile for  without  and with shear cases.}
\begin{table}[!h]
\begin{center}
\caption{Values of the wavenumbers $k_n$ (m$^{-1}$) for the first $5$-modes.}
\label{tab:k_n}
\begin{tabular}{llllll}
\hline
CASE     & $k_1$   &   $k_2$   & $k_3$    & $k_4$   & $k_5$ \\[3pt]
\hline      
(1) without shear  & $1.1\times10^{-4}$  & $4.2\times10^{-4}$ & $5.9\times10^{-4}$
& $9.8\times10^{-4}$ & $10.8\times10^{-4}$ \\
(2) with shear         & $1.1\times10^{-4}$  & $4.4\times10^{-4}$ & $6.2\times10^{-4}$
& $10.4\times10^{-3}$ & $11.6\times10^{-4}$ \\\hline
\end{tabular}
\end{center}
\end{table}

\begin{table}[!h]
\begin{center}
\caption{Estimated values of parameters to reconstruct the Mediterranean density profile using the three-layered models.}
\label{tab:N_n}
\begin{tabular}{llllll}
\hline
CASE     & $N_1$ (s$^{-1}$)   &   $N_2$ (s$^{-1}$)  & $N_3$ (s$^{-1}$)   & $h_1$ (m)   & $h_2$ (m) \\[3pt]
  \hline      
(1) without shear  & $1\times10^{-3}$  & $6.6\times10^{-3}$ & $1.3\times10^{-3}$
& $46$ & $186$ \\
(2) with shear         & $1\times10^{-3}$  & $6.8\times10^{-3}$ & $1.6\times10^{-3}$
& $60$ & $180$ \\\hline
\end{tabular}
\end{center}
\end{table}

{The NRMSE
in the estimation of  density profile is $5.8\%$ for  no--shear case, and $9.8\%$ for with--shear case.
{It follows that  $\delta N/N_{avg} \propto \delta k/k$ where $N_{avg}$ is the average of the layered buoyancy frequency. The error in density estimation $\delta \rho$ is found to be $\delta \rho/\rho \propto N_{avg}^2 \delta N/N_{avg}$.}
{A simple calculation of sensitivity analysis suggests for the case of no-shear and shear the averaged values of $\delta N/N_{avg}$ are respectively  $\delta N/N_{avg} = 2.65 (\delta k/k)$ and $\delta N/N_{avg} = 3.14 (\delta k/k)$. Therefore,  effects of the shear is very small in this particular case. In general, in the oceans $N=\mathcal{O}(10^{-3})$ -- $\mathcal{O}(10^{-1})$ s$^{-1}$, therefore, it can be said that the shear induced modification of $k_n$ weakly affects the density reconstruction, provided the shear is not very strong.} }

\begin{figure}
\centering
\includegraphics[width=0.5\linewidth]{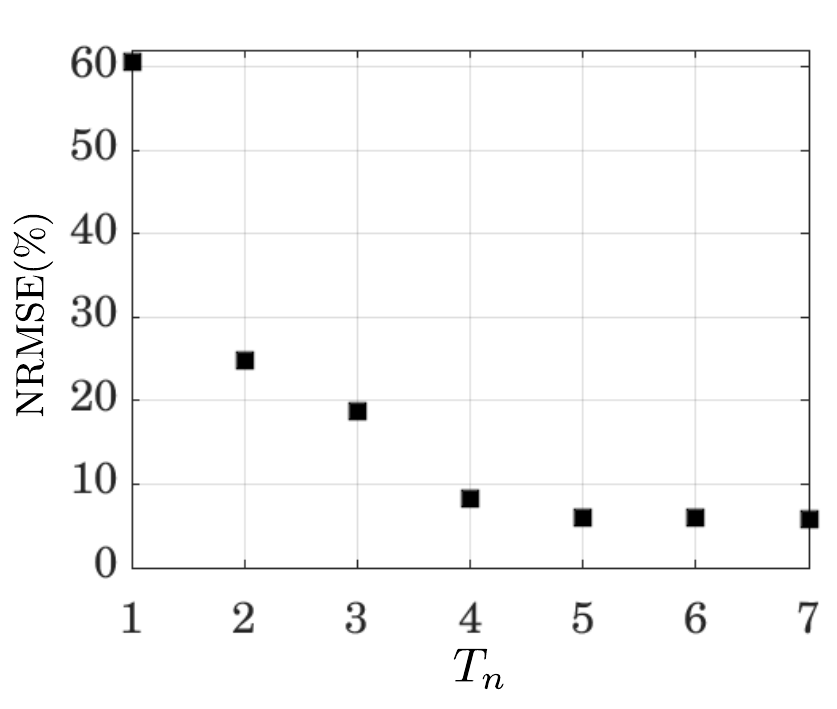}
\caption{NRMSE$(\%)$ versus the first-$n$ modes, $T_n$.  As $n$ increases, the reconstructed error decreases, implying that the reconstructed density profile converges to the actual density profile.}
\label{fig:err}
\end{figure}
{In practice, only the first few modes are available from the satellite altimetry dataset\cite{zhao2012mapping}. Therefore, we have studied the dependence of density reconstruction error as the number of first-$n$ modes, $T_n$ is increased. 
For this particular analysis, we have considered wavenumbers for the `without-shear' case. Fig. \ref{fig:err} shows that the percentage of reconstruction error decreases as the number of modes, $n$ increases. 
The values of $T_n=1$, $3$ and $5$ respectively correspond to one-layer, two-layered and three-layered model. For the case $T_n=2$, we have used Eq. (\ref{eq33}) of the two-layered model with $N_1$ already known from the one-layered model, thereby reducing the number of unknowns to
$2$ (i.e. $N_2$ and $h$), making it solvable. In a similar way, for $T_n=4$ case, we have used Eq. (\ref{eq5}) of the three-layered model with $N_1$ already known from the two-layered model, thereby reducing the number of unknowns to $4$, and reconstruct the density profile
by finding out $N_2$, $N_3$, $h_1$ and $h_2$. Similar procedure has been followed for $T_n=6$ and $T_n=7$ cases. 
{The reconstruction error saturated to $5.8$\% at $T_n=5$. This suggests that the three-layered model can be regarded as an optimal choice for density reconstruction, which is expected since a generic ocean density profile shows a three-layered structure. 
}
We also note that in order to capture the finite thickness of the pycnocline, we need $n > 3$. 
}

\section{Conclusion}
\label{sec:4}
Using  numerical simulations, we show that ocean's free surface carries the information of its mean density profile, and therefore layout the theoretical groundwork  towards its reconstruction. To the best of our knowledge, this is the first work that puts forward an inverse technique towards obtaining or estimating ocean's density profile. The barotropic tides cause the stratified ocean water to move back and forth over the submarine topography, causing the IGW radiation. The IGWs carrying the information of the ocean's mean density profile impinge on the free surface, the signature of which can be observed using the STFT. The wavenumbers (countably infinite in number) constituting this IGW correspond to the tidal frequency in the STFT spectrum. In general, higher the mode number, lower is its amplitude; hence the first few modes of the surface signature (which are  easier to detect in practice) can be used to estimate the density stratification profile. 
The ocean's mean density profile $\bar{\rho}(z)$ has some specific qualities -- it  continuously and monotonically decreases with $z$. Furthermore, the variation of $\bar{\rho}(z)$ is such that it can be broadly divided into three distinct regions. In each of these regions,  $\bar{\rho}(z)$ can be approximated to be linearly varying with $z$ (implying $N$ is constant in each layer). Therefore we construct a `simplified ocean'  with $m$ layers, each with a constant $N$ (the maximum value of $m$ considered here is $3$). The remarkable advantage of this simplification is that a closed form dispersion relation can be obtained. For an $m$-layered flow ($m\geq 2$) we have to find buoyancy frequencies in each layer: $N_1,\,N_2,\,\cdots,\,N_m$, and layer depths: $h_1,\,h_2,\,\cdots,\,h_{m-1}$. It is possible to evaluate these $2m-1$ unknowns by constructing $2m-1$ equations out of the dispersion relation, provided we know the wavenumbers $k_1,\,k_2,\ \cdots,\, k_{2m-1}$ corresponding to the tidal frequency $\omega_0$. Indeed, this information is known from the STFT spectrum of the free surface. 
After reconstructing the simpler profiles, we consider a slightly more  complicated density profile that is representative of the Mediterranean sea. Using the $3$-layered model, we reconstruct the above-mentioned profile with $94.6$\% accuracy. In all these cases,  a Gaussian bottom topography is used. Next we consider a more realistic environment in which rotation is included; moreover bottom topography, density and shear profiles are representative of the Strait of Gibraltar. In the absence of shear, the reconstructed density profile is $94.2$\%  accurate, however accuracy is found to decrease to $90.2$\% in the presence of shear. We note here that, although the complexity of the bottom topography plays little role in altering the wavenumbers of the low-modes and hence, in the density profile reconstruction, the shear can have an important role. In fact,  shear causes loss of coherence of the internal beams; if shear is very strong, it will be very difficult to reconstruct the density profile. One can notice in Fig. \ref{fig:zhao_jpo} the absence of coherent signature of internal tides over a large region near the equatorial Pacific ocean. The cause of this absence has been attributed  to the enhanced dissipation caused by the vertically and horizontally stacked zonal currents with the alternating flow directions \cite{peters1988parameterization}. Fortunately, the effect of shear is small in  most parts of the global ocean, otherwise the coherent signal of mode-$1$ in Fig. \ref{fig:zhao_jpo} wouldn't be present. Hence our technique would be broadly applicable. 
{We have also discussed the dependence of reconstruction error of the density profile on the number of modes since in practice, satellite altimetry dataset provides a finite number of low modes.
}

Finally we discuss the importance and relevance of this work.
There are many theoretical and numerical studies where the stratification of the ocean has been approximated as
layered profiles,  each layer having a constant buoyancy frequency. For example, the the effect of wind-stress on the induced currents in the \emph{top} weakly stratified layer  has often been studied with discrete layers of constant buoyancy frequency, and agrees well with the observation analysis \cite{gill1984behavior,moehlis2001radiation}. The generation of energetic superharmonics using nonlinear self-interaction of the primary linear wave has been explored using simplified discrete constant buoyancy frequency layered models 
and this simplified theoretical model often explains the real ocean
observation \cite{wunsch2017harmonic}. Theoretical studies on the generation of  solitary waves using discrete layers often facilitate understanding of the real ocean scenarios \cite{gerkema2001internal,goullet2008large,wunsch2014simulations}.  Moreover, the knowledge of the location of the ocean pycnocline will be crucially important for designing offshore structures and estimating the impact of the internal solitary waves on those structures\cite{osborne1980internal}.
Furthermore, a serious problem in climate modelling is the accurate simulation of the climatological state of the oceanic density field and this requires a good estimation
of the eddy diffusivity, which is commonly parameterised using the mean buoyancy \cite{cummins1991deep,nilsson2003thermohaline,osborn1980estimates}. Inversion of the mode-1 wavenumbers in Fig. \ref{fig:zhao_jpo} would result in a one-layered global ocean stratification, and if the first five modes are available, the three-layered  global ocean (climatological) density field can be reconstructed with uniform spatial resolution. 
In conclusion, we believe that our proposed technique, in conjunction with the ARGO data, can provide a better estimate of the  global ocean density field in the future. 

\begin{acknowledgments}
A.G. would like to acknowledge the funding support from the Alexander von Humboldt foundation, and the SERB Early Career award  ECR/2016/001493.
\end{acknowledgments}

\bibliography{bibfile}

\begin{thebibliography}{41}%
\makeatletter
\providecommand \@ifxundefined [1]{%
 \@ifx{#1\undefined}
}%
\providecommand \@ifnum [1]{%
 \ifnum #1\expandafter \@firstoftwo
 \else \expandafter \@secondoftwo
 \fi
}%
\providecommand \@ifx [1]{%
 \ifx #1\expandafter \@firstoftwo
 \else \expandafter \@secondoftwo
 \fi
}%
\providecommand \natexlab [1]{#1}%
\providecommand \enquote  [1]{``#1''}%
\providecommand \bibnamefont  [1]{#1}%
\providecommand \bibfnamefont [1]{#1}%
\providecommand \citenamefont [1]{#1}%
\providecommand \href@noop [0]{\@secondoftwo}%
\providecommand \href [0]{\begingroup \@sanitize@url \@href}%
\providecommand \@href[1]{\@@startlink{#1}\@@href}%
\providecommand \@@href[1]{\endgroup#1\@@endlink}%
\providecommand \@sanitize@url [0]{\catcode `\\12\catcode `\$12\catcode
  `\&12\catcode `\#12\catcode `\^12\catcode `\_12\catcode `\%12\relax}%
\providecommand \@@startlink[1]{}%
\providecommand \@@endlink[0]{}%
\providecommand \url  [0]{\begingroup\@sanitize@url \@url }%
\providecommand \@url [1]{\endgroup\@href {#1}{\urlprefix }}%
\providecommand \urlprefix  [0]{URL }%
\providecommand \Eprint [0]{\href }%
\providecommand \doibase [0]{http://dx.doi.org/}%
\providecommand \selectlanguage [0]{\@gobble}%
\providecommand \bibinfo  [0]{\@secondoftwo}%
\providecommand \bibfield  [0]{\@secondoftwo}%
\providecommand \translation [1]{[#1]}%
\providecommand \BibitemOpen [0]{}%
\providecommand \bibitemStop [0]{}%
\providecommand \bibitemNoStop [0]{.\EOS\space}%
\providecommand \EOS [0]{\spacefactor3000\relax}%
\providecommand \BibitemShut  [1]{\csname bibitem#1\endcsname}%
\let\auto@bib@innerbib\@empty
\bibitem [{\citenamefont {Sutherland}(2010)}]{sutherland2010internal}%
  \BibitemOpen
  \bibfield  {author} {\bibinfo {author} {\bibfnamefont {B.~R.}\ \bibnamefont
  {Sutherland}},\ }\href@noop {} {\emph {\bibinfo {title} {Internal gravity
  waves}}}\ (\bibinfo  {publisher} {Cambridge university press},\ \bibinfo
  {year} {2010})\BibitemShut {NoStop}%
\bibitem [{\citenamefont {Cummins}(1991)}]{cummins1991deep}%
  \BibitemOpen
  \bibfield  {author} {\bibinfo {author} {\bibfnamefont {P.~F.}\ \bibnamefont
  {Cummins}},\ }\href@noop {} {\bibfield  {journal} {\bibinfo  {journal}
  {Atmos.-Ocean}\ }\textbf {\bibinfo {volume} {29}},\ \bibinfo {pages} {563}
  (\bibinfo {year} {1991})}\BibitemShut {NoStop}%
\bibitem [{\citenamefont {Doostmohammadi}, \citenamefont {Stocker},\ and\
  \citenamefont {Ardekani}(2012)}]{doostmohammadi2012low}%
  \BibitemOpen
  \bibfield  {author} {\bibinfo {author} {\bibfnamefont {A.}~\bibnamefont
  {Doostmohammadi}}, \bibinfo {author} {\bibfnamefont {R.}~\bibnamefont
  {Stocker}}, \ and\ \bibinfo {author} {\bibfnamefont {A.~M.}\ \bibnamefont
  {Ardekani}},\ }\href@noop {} {\bibfield  {journal} {\bibinfo  {journal}
  {Proc. Nat. Acad. Sci.}\ } (\bibinfo {year} {2012})}\BibitemShut {NoStop}%
\bibitem [{\citenamefont {Sherman}\ \emph {et~al.}(1998)\citenamefont
  {Sherman}, \citenamefont {Webster}, \citenamefont {Jones},\ and\
  \citenamefont {Oliver}}]{sherman1998transitions}%
  \BibitemOpen
  \bibfield  {author} {\bibinfo {author} {\bibfnamefont {B.~S.}\ \bibnamefont
  {Sherman}}, \bibinfo {author} {\bibfnamefont {I.~T.}\ \bibnamefont
  {Webster}}, \bibinfo {author} {\bibfnamefont {G.~J.}\ \bibnamefont {Jones}},
  \ and\ \bibinfo {author} {\bibfnamefont {R.~L.}\ \bibnamefont {Oliver}},\
  }\href@noop {} {\bibfield  {journal} {\bibinfo  {journal} {Limnol.
  Oceanogr.}\ }\textbf {\bibinfo {volume} {43}},\ \bibinfo {pages} {1902}
  (\bibinfo {year} {1998})}\BibitemShut {NoStop}%
\bibitem [{\citenamefont {Capotondi}\ \emph {et~al.}(2012)\citenamefont
  {Capotondi}, \citenamefont {Alexander}, \citenamefont {Bond}, \citenamefont
  {Curchitser},\ and\ \citenamefont {Scott}}]{capotondi2012enhanced}%
  \BibitemOpen
  \bibfield  {author} {\bibinfo {author} {\bibfnamefont {A.}~\bibnamefont
  {Capotondi}}, \bibinfo {author} {\bibfnamefont {M.~A.}\ \bibnamefont
  {Alexander}}, \bibinfo {author} {\bibfnamefont {N.~A.}\ \bibnamefont {Bond}},
  \bibinfo {author} {\bibfnamefont {E.~N.}\ \bibnamefont {Curchitser}}, \ and\
  \bibinfo {author} {\bibfnamefont {J.~D.}\ \bibnamefont {Scott}},\ }\href@noop
  {} {\bibfield  {journal} {\bibinfo  {journal} {J. Geophys. Res. Oceans.}\
  }\textbf {\bibinfo {volume} {117}} (\bibinfo {year} {2012})}\BibitemShut
  {NoStop}%
\bibitem [{\citenamefont {Bradshaw}\ and\ \citenamefont
  {Schleicher}(1980)}]{bradshaw1980electrical}%
  \BibitemOpen
  \bibfield  {author} {\bibinfo {author} {\bibfnamefont {A.}~\bibnamefont
  {Bradshaw}}\ and\ \bibinfo {author} {\bibfnamefont {K.}~\bibnamefont
  {Schleicher}},\ }\href@noop {} {\bibfield  {journal} {\bibinfo  {journal}
  {IEEE J. Oceanic Eng.}\ }\textbf {\bibinfo {volume} {5}},\ \bibinfo {pages}
  {50} (\bibinfo {year} {1980})}\BibitemShut {NoStop}%
\bibitem [{\citenamefont {Wunsch}(1975)}]{wunsch1975internal}%
  \BibitemOpen
  \bibfield  {author} {\bibinfo {author} {\bibfnamefont {C.}~\bibnamefont
  {Wunsch}},\ }\href@noop {} {\bibfield  {journal} {\bibinfo  {journal} {Rev.
  Geophys.}\ }\textbf {\bibinfo {volume} {13}},\ \bibinfo {pages} {167}
  (\bibinfo {year} {1975})}\BibitemShut {NoStop}%
\bibitem [{\citenamefont {Zhao}(2016)}]{zhao2016internal}%
  \BibitemOpen
  \bibfield  {author} {\bibinfo {author} {\bibfnamefont {Z.}~\bibnamefont
  {Zhao}},\ }\href@noop {} {\bibfield  {journal} {\bibinfo  {journal}
  {Geophysical Research Letters}\ }\textbf {\bibinfo {volume} {43}},\ \bibinfo
  {pages} {9157} (\bibinfo {year} {2016})}\BibitemShut {NoStop}%
\bibitem [{\citenamefont {Alford}\ \emph {et~al.}(2016)\citenamefont {Alford},
  \citenamefont {MacKinnon}, \citenamefont {Simmons},\ and\ \citenamefont
  {Nash}}]{alford2016near}%
  \BibitemOpen
  \bibfield  {author} {\bibinfo {author} {\bibfnamefont {M.~H.}\ \bibnamefont
  {Alford}}, \bibinfo {author} {\bibfnamefont {J.~A.}\ \bibnamefont
  {MacKinnon}}, \bibinfo {author} {\bibfnamefont {H.~L.}\ \bibnamefont
  {Simmons}}, \ and\ \bibinfo {author} {\bibfnamefont {J.~D.}\ \bibnamefont
  {Nash}},\ }\href@noop {} {\bibfield  {journal} {\bibinfo  {journal} {Annu.
  Rev. Mar. Sci.}\ }\textbf {\bibinfo {volume} {8}},\ \bibinfo {pages} {95}
  (\bibinfo {year} {2016})}\BibitemShut {NoStop}%
\bibitem [{\citenamefont {D'Asaro}(1989)}]{d1989decay}%
  \BibitemOpen
  \bibfield  {author} {\bibinfo {author} {\bibfnamefont {E.~A.}\ \bibnamefont
  {D'Asaro}},\ }\href@noop {} {\bibfield  {journal} {\bibinfo  {journal} {J.
  Geophys. Res. Oceans.}\ }\textbf {\bibinfo {volume} {94}},\ \bibinfo {pages}
  {2045} (\bibinfo {year} {1989})}\BibitemShut {NoStop}%
\bibitem [{\citenamefont {Garrett}(2001)}]{garrett2001near}%
  \BibitemOpen
  \bibfield  {author} {\bibinfo {author} {\bibfnamefont {C.}~\bibnamefont
  {Garrett}},\ }\href@noop {} {\bibfield  {journal} {\bibinfo  {journal} {J.
  Phys. Oceanogr.}\ }\textbf {\bibinfo {volume} {31}},\ \bibinfo {pages} {962}
  (\bibinfo {year} {2001})}\BibitemShut {NoStop}%
\bibitem [{\citenamefont {Zhao}\ \emph {et~al.}(2016)\citenamefont {Zhao},
  \citenamefont {Alford}, \citenamefont {Girton}, \citenamefont {Rainville},\
  and\ \citenamefont {Simmons}}]{zhao2016global}%
  \BibitemOpen
  \bibfield  {author} {\bibinfo {author} {\bibfnamefont {Z.}~\bibnamefont
  {Zhao}}, \bibinfo {author} {\bibfnamefont {M.~H.}\ \bibnamefont {Alford}},
  \bibinfo {author} {\bibfnamefont {J.~B.}\ \bibnamefont {Girton}}, \bibinfo
  {author} {\bibfnamefont {L.}~\bibnamefont {Rainville}}, \ and\ \bibinfo
  {author} {\bibfnamefont {H.~L.}\ \bibnamefont {Simmons}},\ }\href@noop {}
  {\bibfield  {journal} {\bibinfo  {journal} {J. Phys. Oceanogr.}\ }\textbf
  {\bibinfo {volume} {46}},\ \bibinfo {pages} {1657} (\bibinfo {year}
  {2016})}\BibitemShut {NoStop}%
\bibitem [{\citenamefont {St.~Laurent}\ and\ \citenamefont
  {Garrett}(2002)}]{st2002role}%
  \BibitemOpen
  \bibfield  {author} {\bibinfo {author} {\bibfnamefont {L.}~\bibnamefont
  {St.~Laurent}}\ and\ \bibinfo {author} {\bibfnamefont {C.}~\bibnamefont
  {Garrett}},\ }\href@noop {} {\bibfield  {journal} {\bibinfo  {journal} {J.
  Phys. Oceanogr.}\ }\textbf {\bibinfo {volume} {32}},\ \bibinfo {pages} {2882}
  (\bibinfo {year} {2002})}\BibitemShut {NoStop}%
\bibitem [{\citenamefont {Sarkar}\ and\ \citenamefont
  {Scotti}(2017)}]{sarkar2017topographic}%
  \BibitemOpen
  \bibfield  {author} {\bibinfo {author} {\bibfnamefont {S.}~\bibnamefont
  {Sarkar}}\ and\ \bibinfo {author} {\bibfnamefont {A.}~\bibnamefont
  {Scotti}},\ }\href@noop {} {\bibfield  {journal} {\bibinfo  {journal} {Annu.
  Rev. Fluid Mech.}\ }\textbf {\bibinfo {volume} {49}},\ \bibinfo {pages} {195}
  (\bibinfo {year} {2017})}\BibitemShut {NoStop}%
\bibitem [{\citenamefont {Ferrari}\ and\ \citenamefont
  {Wunsch}(2009)}]{ferrari2009ocean}%
  \BibitemOpen
  \bibfield  {author} {\bibinfo {author} {\bibfnamefont {R.}~\bibnamefont
  {Ferrari}}\ and\ \bibinfo {author} {\bibfnamefont {C.}~\bibnamefont
  {Wunsch}},\ }\href@noop {} {\bibfield  {journal} {\bibinfo  {journal} {Annu.
  Rev. Fluid Mech.}\ }\textbf {\bibinfo {volume} {41}} (\bibinfo {year}
  {2009})}\BibitemShut {NoStop}%
\bibitem [{\citenamefont {Ray}\ and\ \citenamefont
  {Mitchum}(1996)}]{ray1996surface}%
  \BibitemOpen
  \bibfield  {author} {\bibinfo {author} {\bibfnamefont {R.~D.}\ \bibnamefont
  {Ray}}\ and\ \bibinfo {author} {\bibfnamefont {G.~T.}\ \bibnamefont
  {Mitchum}},\ }\href@noop {} {\bibfield  {journal} {\bibinfo  {journal}
  {Geophys. Res. Lett.}\ }\textbf {\bibinfo {volume} {23}},\ \bibinfo {pages}
  {2101} (\bibinfo {year} {1996})}\BibitemShut {NoStop}%
\bibitem [{\citenamefont {Ray}\ and\ \citenamefont {Zaron}(2011)}]{ray2011non}%
  \BibitemOpen
  \bibfield  {author} {\bibinfo {author} {\bibfnamefont {R.~D.}\ \bibnamefont
  {Ray}}\ and\ \bibinfo {author} {\bibfnamefont {E.~D.}\ \bibnamefont
  {Zaron}},\ }\href@noop {} {\bibfield  {journal} {\bibinfo  {journal}
  {Geophys. Res. Lett.}\ }\textbf {\bibinfo {volume} {38}} (\bibinfo {year}
  {2011})}\BibitemShut {NoStop}%
\bibitem [{\citenamefont {Zhao}, \citenamefont {Alford},\ and\ \citenamefont
  {Girton}(2012)}]{zhao2012mapping}%
  \BibitemOpen
  \bibfield  {author} {\bibinfo {author} {\bibfnamefont {Z.}~\bibnamefont
  {Zhao}}, \bibinfo {author} {\bibfnamefont {M.~H.}\ \bibnamefont {Alford}}, \
  and\ \bibinfo {author} {\bibfnamefont {J.~B.}\ \bibnamefont {Girton}},\
  }\href@noop {} {\bibfield  {journal} {\bibinfo  {journal} {Oceanography}\
  }\textbf {\bibinfo {volume} {25}},\ \bibinfo {pages} {42} (\bibinfo {year}
  {2012})}\BibitemShut {NoStop}%
\bibitem [{\citenamefont {Gerkema}(2001)}]{gerkema2001internal}%
  \BibitemOpen
  \bibfield  {author} {\bibinfo {author} {\bibfnamefont {T.}~\bibnamefont
  {Gerkema}},\ }\href@noop {} {\bibfield  {journal} {\bibinfo  {journal} {J.
  Mar. Res.}\ }\textbf {\bibinfo {volume} {59}},\ \bibinfo {pages} {227}
  (\bibinfo {year} {2001})}\BibitemShut {NoStop}%
\bibitem [{\citenamefont {Gerkema}\ and\ \citenamefont
  {Zimmerman}(2008)}]{gerkema2008introduction}%
  \BibitemOpen
  \bibfield  {author} {\bibinfo {author} {\bibfnamefont {T.}~\bibnamefont
  {Gerkema}}\ and\ \bibinfo {author} {\bibfnamefont {J.~T.~F.}\ \bibnamefont
  {Zimmerman}},\ }\href@noop {} {\bibfield  {journal} {\bibinfo  {journal}
  {Lecture Notes, Royal NIOZ, Texel}\ }\textbf {\bibinfo {volume} {207}}
  (\bibinfo {year} {2008})}\BibitemShut {NoStop}%
\bibitem [{\citenamefont {Verma}(2018)}]{kumar2018physics}%
  \BibitemOpen
  \bibfield  {author} {\bibinfo {author} {\bibfnamefont {M.~K.}\ \bibnamefont
  {Verma}},\ }\href@noop {} {\emph {\bibinfo {title} {Physics Of Buoyant Flows:
  From Instabilities To Turbulence}}}\ (\bibinfo  {publisher} {World
  Scientific},\ \bibinfo {year} {2018})\BibitemShut {NoStop}%
\bibitem [{\citenamefont {Wunsch}(2013)}]{wunsch2013baroclinic}%
  \BibitemOpen
  \bibfield  {author} {\bibinfo {author} {\bibfnamefont {C.}~\bibnamefont
  {Wunsch}},\ }\href@noop {} {\bibfield  {journal} {\bibinfo  {journal}
  {Journal of Atmospheric and Oceanic Technology}\ }\textbf {\bibinfo {volume}
  {30}},\ \bibinfo {pages} {140} (\bibinfo {year} {2013})}\BibitemShut
  {NoStop}%
\bibitem [{\citenamefont {Turner}(1979)}]{turner1979buoyancy}%
  \BibitemOpen
  \bibfield  {author} {\bibinfo {author} {\bibfnamefont {J.~S.}\ \bibnamefont
  {Turner}},\ }\href@noop {} {\emph {\bibinfo {title} {Buoyancy effects in
  fluids}}}\ (\bibinfo  {publisher} {Cambridge University Press},\ \bibinfo
  {year} {1979})\BibitemShut {NoStop}%
\bibitem [{\citenamefont {Zettl}(2010)}]{zettl2010sturm}%
  \BibitemOpen
  \bibfield  {author} {\bibinfo {author} {\bibfnamefont {A.}~\bibnamefont
  {Zettl}},\ }\href@noop {} {\emph {\bibinfo {title} {Sturm-{L}iouville
  theory}}},\ \bibinfo {number} {121}\ (\bibinfo  {publisher} {American
  Mathematical Soc.},\ \bibinfo {year} {2010})\BibitemShut {NoStop}%
\bibitem [{\citenamefont {Dimas}\ and\ \citenamefont
  {Triantafyllou}(1994)}]{dimas1994nonlinear}%
  \BibitemOpen
  \bibfield  {author} {\bibinfo {author} {\bibfnamefont {A.~A.}\ \bibnamefont
  {Dimas}}\ and\ \bibinfo {author} {\bibfnamefont {G.~S.}\ \bibnamefont
  {Triantafyllou}},\ }\href@noop {} {\bibfield  {journal} {\bibinfo  {journal}
  {J. Fluid Mech.}\ }\textbf {\bibinfo {volume} {260}},\ \bibinfo {pages} {211}
  (\bibinfo {year} {1994})}\BibitemShut {NoStop}%
\bibitem [{\citenamefont {Burns}\ \emph {et~al.}(2019)\citenamefont {Burns},
  \citenamefont {Vasil}, \citenamefont {Oishi}, \citenamefont {Lecoanet},\ and\
  \citenamefont {Brown}}]{burns2019dedalus}%
  \BibitemOpen
  \bibfield  {author} {\bibinfo {author} {\bibfnamefont {K.~J.}\ \bibnamefont
  {Burns}}, \bibinfo {author} {\bibfnamefont {G.~M.}\ \bibnamefont {Vasil}},
  \bibinfo {author} {\bibfnamefont {J.~S.}\ \bibnamefont {Oishi}}, \bibinfo
  {author} {\bibfnamefont {D.}~\bibnamefont {Lecoanet}}, \ and\ \bibinfo
  {author} {\bibfnamefont {B.~P.}\ \bibnamefont {Brown}},\ }\href@noop {}
  {\bibfield  {journal} {\bibinfo  {journal} {arXiv preprint arXiv:1905.10388}\
  } (\bibinfo {year} {2019})}\BibitemShut {NoStop}%
\bibitem [{\citenamefont {Send}\ and\ \citenamefont
  {Baschek}(2001)}]{send2001intensive}%
  \BibitemOpen
  \bibfield  {author} {\bibinfo {author} {\bibfnamefont {U.}~\bibnamefont
  {Send}}\ and\ \bibinfo {author} {\bibfnamefont {B.}~\bibnamefont {Baschek}},\
  }\href@noop {} {\bibfield  {journal} {\bibinfo  {journal} {J. Geophys. Res.
  Oceans.}\ }\textbf {\bibinfo {volume} {106}},\ \bibinfo {pages} {31017}
  (\bibinfo {year} {2001})}\BibitemShut {NoStop}%
\bibitem [{\citenamefont {Weatherall}\ \emph {et~al.}(2015)\citenamefont
  {Weatherall}, \citenamefont {Marks}, \citenamefont {Jakobsson}, \citenamefont
  {Schmitt}, \citenamefont {Tani}, \citenamefont {Arndt}, \citenamefont
  {Rovere}, \citenamefont {Chayes}, \citenamefont {Ferrini},\ and\
  \citenamefont {Wigley}}]{weatherall2015new}%
  \BibitemOpen
  \bibfield  {author} {\bibinfo {author} {\bibfnamefont {P.}~\bibnamefont
  {Weatherall}}, \bibinfo {author} {\bibfnamefont {K.~M.}\ \bibnamefont
  {Marks}}, \bibinfo {author} {\bibfnamefont {M.}~\bibnamefont {Jakobsson}},
  \bibinfo {author} {\bibfnamefont {T.}~\bibnamefont {Schmitt}}, \bibinfo
  {author} {\bibfnamefont {S.}~\bibnamefont {Tani}}, \bibinfo {author}
  {\bibfnamefont {J.~E.}\ \bibnamefont {Arndt}}, \bibinfo {author}
  {\bibfnamefont {M.}~\bibnamefont {Rovere}}, \bibinfo {author} {\bibfnamefont
  {D.}~\bibnamefont {Chayes}}, \bibinfo {author} {\bibfnamefont
  {V.}~\bibnamefont {Ferrini}}, \ and\ \bibinfo {author} {\bibfnamefont
  {R.}~\bibnamefont {Wigley}},\ }\href@noop {} {\bibfield  {journal} {\bibinfo
  {journal} {Earth Space Sci.}\ }\textbf {\bibinfo {volume} {2}},\ \bibinfo
  {pages} {331} (\bibinfo {year} {2015})}\BibitemShut {NoStop}%
\bibitem [{\citenamefont {Marshall}\ \emph {et~al.}(1997)\citenamefont
  {Marshall}, \citenamefont {Adcroft}, \citenamefont {Hill}, \citenamefont
  {Perelman},\ and\ \citenamefont {Heisey}}]{marshall1997finite}%
  \BibitemOpen
  \bibfield  {author} {\bibinfo {author} {\bibfnamefont {J.}~\bibnamefont
  {Marshall}}, \bibinfo {author} {\bibfnamefont {A.}~\bibnamefont {Adcroft}},
  \bibinfo {author} {\bibfnamefont {C.}~\bibnamefont {Hill}}, \bibinfo {author}
  {\bibfnamefont {L.}~\bibnamefont {Perelman}}, \ and\ \bibinfo {author}
  {\bibfnamefont {C.}~\bibnamefont {Heisey}},\ }\href@noop {} {\bibfield
  {journal} {\bibinfo  {journal} {J. Geophys. Res. Oceans}\ }\textbf {\bibinfo
  {volume} {102}},\ \bibinfo {pages} {5753} (\bibinfo {year}
  {1997})}\BibitemShut {NoStop}%
\bibitem [{\citenamefont {Adcroft}, \citenamefont {Hill},\ and\ \citenamefont
  {Marshall}(1997)}]{adcroft1997representation}%
  \BibitemOpen
  \bibfield  {author} {\bibinfo {author} {\bibfnamefont {A.}~\bibnamefont
  {Adcroft}}, \bibinfo {author} {\bibfnamefont {C.}~\bibnamefont {Hill}}, \
  and\ \bibinfo {author} {\bibfnamefont {J.}~\bibnamefont {Marshall}},\
  }\href@noop {} {\bibfield  {journal} {\bibinfo  {journal} {Mon. Weather
  Rev.}\ }\textbf {\bibinfo {volume} {125}},\ \bibinfo {pages} {2293} (\bibinfo
  {year} {1997})}\BibitemShut {NoStop}%
\bibitem [{\citenamefont {Izquierdo}\ and\ \citenamefont
  {Mikolajewicz}(2019)}]{izquierdo2019role}%
  \BibitemOpen
  \bibfield  {author} {\bibinfo {author} {\bibfnamefont {A.}~\bibnamefont
  {Izquierdo}}\ and\ \bibinfo {author} {\bibfnamefont {U.}~\bibnamefont
  {Mikolajewicz}},\ }\href@noop {} {\bibfield  {journal} {\bibinfo  {journal}
  {Ocean Model.}\ }\textbf {\bibinfo {volume} {133}},\ \bibinfo {pages} {27}
  (\bibinfo {year} {2019})}\BibitemShut {NoStop}%
\bibitem [{\citenamefont {Drazin}\ and\ \citenamefont {Reid}(2004)}]{draz1982}%
  \BibitemOpen
  \bibfield  {author} {\bibinfo {author} {\bibfnamefont {P.~G.}\ \bibnamefont
  {Drazin}}\ and\ \bibinfo {author} {\bibfnamefont {W.~H.}\ \bibnamefont
  {Reid}},\ }\href@noop {} {\emph {\bibinfo {title} {{H}ydrodynamic
  {S}tability}}},\ \bibinfo {edition} {2nd}\ ed.\ (\bibinfo  {publisher}
  {Cambridge University Press},\ \bibinfo {year} {2004})\BibitemShut {NoStop}%
\bibitem [{\citenamefont {Peters}, \citenamefont {Gregg},\ and\ \citenamefont
  {Toole}(1988)}]{peters1988parameterization}%
  \BibitemOpen
  \bibfield  {author} {\bibinfo {author} {\bibfnamefont {H.}~\bibnamefont
  {Peters}}, \bibinfo {author} {\bibfnamefont {M.~C.}\ \bibnamefont {Gregg}}, \
  and\ \bibinfo {author} {\bibfnamefont {J.~M.}\ \bibnamefont {Toole}},\
  }\href@noop {} {\bibfield  {journal} {\bibinfo  {journal} {J. Geophys. Res.
  Oceans.}\ }\textbf {\bibinfo {volume} {93}},\ \bibinfo {pages} {1199}
  (\bibinfo {year} {1988})}\BibitemShut {NoStop}%
\bibitem [{\citenamefont {Gill}(1984)}]{gill1984behavior}%
  \BibitemOpen
  \bibfield  {author} {\bibinfo {author} {\bibfnamefont {A.~E.}\ \bibnamefont
  {Gill}},\ }\href@noop {} {\bibfield  {journal} {\bibinfo  {journal} {J. Phys.
  Oceanogr.}\ }\textbf {\bibinfo {volume} {14}},\ \bibinfo {pages} {1129}
  (\bibinfo {year} {1984})}\BibitemShut {NoStop}%
\bibitem [{\citenamefont {Moehlis}\ and\ \citenamefont
  {Llewellyn~Smith}(2001)}]{moehlis2001radiation}%
  \BibitemOpen
  \bibfield  {author} {\bibinfo {author} {\bibfnamefont {J.}~\bibnamefont
  {Moehlis}}\ and\ \bibinfo {author} {\bibfnamefont {S.~G.}\ \bibnamefont
  {Llewellyn~Smith}},\ }\href@noop {} {\bibfield  {journal} {\bibinfo
  {journal} {J. Phys. Oceanogr.}\ }\textbf {\bibinfo {volume} {31}},\ \bibinfo
  {pages} {1550} (\bibinfo {year} {2001})}\BibitemShut {NoStop}%
\bibitem [{\citenamefont {Wunsch}(2017)}]{wunsch2017harmonic}%
  \BibitemOpen
  \bibfield  {author} {\bibinfo {author} {\bibfnamefont {S.}~\bibnamefont
  {Wunsch}},\ }\href@noop {} {\bibfield  {journal} {\bibinfo  {journal} {J.
  Fluid Mech.}\ }\textbf {\bibinfo {volume} {828}},\ \bibinfo {pages} {630}
  (\bibinfo {year} {2017})}\BibitemShut {NoStop}%
\bibitem [{\citenamefont {Goullet}\ and\ \citenamefont
  {Choi}(2008)}]{goullet2008large}%
  \BibitemOpen
  \bibfield  {author} {\bibinfo {author} {\bibfnamefont {A.}~\bibnamefont
  {Goullet}}\ and\ \bibinfo {author} {\bibfnamefont {W.}~\bibnamefont {Choi}},\
  }\href@noop {} {\bibfield  {journal} {\bibinfo  {journal} {Phys. Fluids}\
  }\textbf {\bibinfo {volume} {20}},\ \bibinfo {pages} {096601} (\bibinfo
  {year} {2008})}\BibitemShut {NoStop}%
\bibitem [{\citenamefont {Wunsch}\ \emph {et~al.}(2014)\citenamefont {Wunsch},
  \citenamefont {Ku}, \citenamefont {Delwiche},\ and\ \citenamefont
  {Awadallah}}]{wunsch2014simulations}%
  \BibitemOpen
  \bibfield  {author} {\bibinfo {author} {\bibfnamefont {S.}~\bibnamefont
  {Wunsch}}, \bibinfo {author} {\bibfnamefont {H.}~\bibnamefont {Ku}}, \bibinfo
  {author} {\bibfnamefont {I.}~\bibnamefont {Delwiche}}, \ and\ \bibinfo
  {author} {\bibfnamefont {R.}~\bibnamefont {Awadallah}},\ }\href@noop {}
  {\bibfield  {journal} {\bibinfo  {journal} {Nonlinear Process. Geophys.}\
  }\textbf {\bibinfo {volume} {21}},\ \bibinfo {pages} {855} (\bibinfo {year}
  {2014})}\BibitemShut {NoStop}%
\bibitem [{\citenamefont {Osborne}\ and\ \citenamefont
  {Burch}(1980)}]{osborne1980internal}%
  \BibitemOpen
  \bibfield  {author} {\bibinfo {author} {\bibfnamefont {A.~R.}\ \bibnamefont
  {Osborne}}\ and\ \bibinfo {author} {\bibfnamefont {T.~L.}\ \bibnamefont
  {Burch}},\ }\href@noop {} {\bibfield  {journal} {\bibinfo  {journal}
  {Science}\ }\textbf {\bibinfo {volume} {208}},\ \bibinfo {pages} {451}
  (\bibinfo {year} {1980})}\BibitemShut {NoStop}%
\bibitem [{\citenamefont {Nilsson}, \citenamefont {Brostr{\"o}m},\ and\
  \citenamefont {Walin}(2003)}]{nilsson2003thermohaline}%
  \BibitemOpen
  \bibfield  {author} {\bibinfo {author} {\bibfnamefont {J.}~\bibnamefont
  {Nilsson}}, \bibinfo {author} {\bibfnamefont {G.}~\bibnamefont
  {Brostr{\"o}m}}, \ and\ \bibinfo {author} {\bibfnamefont {G.}~\bibnamefont
  {Walin}},\ }\href@noop {} {\bibfield  {journal} {\bibinfo  {journal} {J.
  Phys. Oceanogr.}\ }\textbf {\bibinfo {volume} {33}},\ \bibinfo {pages} {2781}
  (\bibinfo {year} {2003})}\BibitemShut {NoStop}%
\bibitem [{\citenamefont {Osborn}(1980)}]{osborn1980estimates}%
  \BibitemOpen
  \bibfield  {author} {\bibinfo {author} {\bibfnamefont {T.}~\bibnamefont
  {Osborn}},\ }\href@noop {} {\bibfield  {journal} {\bibinfo  {journal} {J.
  Phys. Oceanogr.}\ }\textbf {\bibinfo {volume} {10}},\ \bibinfo {pages} {83}
  (\bibinfo {year} {1980})}\BibitemShut {NoStop}%
\end{thebibliography}%
\end{document}